\documentclass[english, preprint, authoryear]{elsarticle}
\usepackage[T1]{fontenc}
\usepackage[latin9]{inputenc}
\usepackage{textcomp}
\usepackage{url}
\usepackage{amsmath}
\usepackage{graphicx}

\makeatletter

\providecommand{\tabularnewline}{\\}
\newcommand{\lyxdot}{.}

\usepackage{mmap} 

\makeatother

\usepackage{babel}
\begin{document}

\title{Biological control of the chestnut gall wasp with \emph{T. sinensis}:
a mathematical model.}

\author[LE]{Francesco Paparella\corref{cor1}}

\ead{francesco.paparella@nyu.edu}

\author[TO]{Chiara Ferracini\corref{cor2}}

\ead{chiara.ferracini@unito.it}

\author[TO]{Alessandro Portaluri}

\author[MI]{Alberto Manzo}

\author[TO]{Alberto Alma}

\address[LE]{Division of Sciences - New York University Abu Dhabi - UAE}

\address[TO]{Dep. of Agricultural, Forest and Food Sciences - University of Torino
- Italy}

\address[MI]{Ministry of Agriculture, Food and Forestry - Italy}

\cortext[cor1]{Principal corresponding author. Permanent address: Dip. di Matematica
\& Fisica, Università del Salento, Lecce, Italy.}

\cortext[cor2]{Corresponding author.}
\begin{abstract}
The Asian chestnut gall wasp \emph{Dryocosmus kuriphilus}, native
of China, has become a pest when it appeared in Japan, Korea, and
the United States. In Europe it was first found in Italy, in 2002.
In 1982 the host-specific parasitoid \emph{Torymus sinensis} was introduced
in Japan, in an attempt to achieve a biological control of the pest.
After an apparent initial success, the two species seem to have locked
in predator-prey cycles of decadal length. We have developed a spatially
explicit mathematical model that describes the seasonal time evolution
of the adult insect populations, and the competition for finding egg
deposition sites. In a spatially homogeneous situation the model reduces
to an iterated map for the egg density of the two species. While the
map would suggest, for realistic parameters, that both species should
become locally extinct (somewhat corroborating the hypothesis of biological
control), the full model, for the same parameters, shows that the
introduction of \emph{T. sinensis} sparks a traveling wave of the
parasitoid population that destroys the pest on its passage. Depending
on the value of the diffusion coefficients of the two species, the
pest can later be able to re-colonize the empty area left behind the
wave. When this occurs the two populations do not seem to attain a
state of spatial homogeneity, but produce an ever-changing pattern
of traveling waves. 
\end{abstract}
\maketitle

\section{Introduction}

Since its first report in 2002 the Asian chestnut gall wasp \emph{Dryocosmus
kuriphilus} Yasumatsu (Hymenoptera: Cynipidae) is affecting many chestnut
ecosystems in Europe and its range is continuously expanding. Native
of China, it established as a pest in the mid 20th century in several
countries, being reported in Japan (1941) \citep{Moriya03}, in Korea
(1958) \citep{Cho&Lee63}, in the United States (1974) \citep{Rieske07}
in Nepal (1999) \citep{Abe07}, and in Canada (2012) \citep{Huber12}. 

In Europe, \emph{D. kuriphilus} was first found in Italy and reported
only in 2002 \citep{Brussino02}. It was added to the European Plant
Protection Organization (EPPO) A2 Action list \citep{Eppo05} in 2003.
Despite the European Commission Decision 2006/464/EC of 27 June 2006
to put into place provisional emergency measures to prevent the introduction
into and the spread within the community of \emph{D. kuriphilus},
the pest is now widely distributed in Italy and has become established
in many other European countries including Austria (2013), Croatia
(2010), Czech Republic and Slovakia (2012), France (2005), Germany
(2013), Hungary (2013), Portugal (2014), Slovenia (2005), Spain (2012),
Switzerland (2009), Turkey (2014) and the United Kingdom (2015) \citep{EFSA10,Eppo13,Eppo15}.
In the Netherlands it was accidentally imported through nursery trees
(2010) and then promptly detected and eradicated by destroying the
few affected plants \citep{NPPO13}, but recently a new outbreak has
been detected close to the German border \citep{Eppo15b}. Since \emph{D.
kuriphilus} has shown its ability to spread rapidly and is successfully
established in several countries, further establishment is likely
in Europe anywhere there is availability of the host plants \emph{Castanea}
spp. \citep{EFSA10}. 

The chestnut gall wasp is a univoltine and thelytokous species \citep{Moriya89},
and lays eggs in buds during summer. The hatched larvae induce the
formation of greenish-red galls, which develop at the time of budburst
in the following early spring on new shoots \citep{Otake80}, suppressing
shoot elongation and causing twig dieback. Severe reduction of fruiting
with yield losses due to insect attacks have been estimated to reach
between 65\% and 85\% in northern Italy \citep{Bosio13,Battisti14}.
However, no evidence was found to confirm tree mortality. A gradual
reduction in vigor in the longer term is the likely consequence of
annual infestation by the gall wasp, causing a gradual reduction in
biomass \citep{EFSA10}. 

Early attempts of biological control of the pest were performed in
Japan \citep{Murakami77,Murakami81} and in the USA \citep{Rieske07}
by the introduction of \emph{Torymus sinensis} Kamijo (Hymenoptera:
Torymidae), a chinese parasitoid described by \citet{Kamijo82}. In
its native environment it is only one among many species of natural
parasitoids of \emph{D. kuriphilus} \citep{Murakami80}, but it appears
to be very well synchronized with the life cycle of the pest, making
it a strong candidate as a biological control agent \citep{Murakami81}.
In addition, outside China, it was believed to be host-specific, although
its host range was never studied or tested in detail \citep{Murakami77,Gibbs11}.
Recently, a large-scale survey in northern Italy found a few specimens
of \emph{T. sinensis} emerging from oak galls of the non-target host
\emph{Biorhiza pallida} Olivier. All evidence, however, still suggests
that \emph{D. kuriphilus} is by far the preferred host, and parasitism
of other species occurs only exceptionally, possibly as a response
to scarcity of its primary host \citep{Ferracini15a}. 

\emph{T. sinensis} reproduces sexually, and by arrhenotokous parthenogenesis
if there is lack of mating. It is reported as univoltine, like its
host. However, recent preliminary investigations highlighted that
a very small fraction of the insect population may undergo a prolonged
diapause extended for 12 months, mainly as late instar larva \citep{Ferracini15b}.
After emergence, which takes place in early spring, and mating, the
female lays eggs inside the larval chamber of newly formed galls,
one egg per host larva. Although in controlled conditions occasional
multiple eggs per host larva have been reported by an early study
\citep{Piao92}, we have never found more than one egg per host larva
while dissecting galls collected in the field. After hatching, the
larva feeds ectoparasitically on the host larva, and it pupates in
the host larval chamber during winter. 

\emph{T. sinensis} was introduced in Japan from China \citep{Murakami77,Murakami80,Moriya03}.
After its release, it dispersed successfully alongside expanding \emph{D.
kuriphilus} populations. In Japan \emph{D. kuriphilus} may also be
subject to varying levels of parasitism from native insects, most
notably \emph{Torymus beneficus} Yasumatsu \& Kamijo and several species
of the genus \emph{Eupelmus} \citep{Murakami95,Moriya03} that, however,
are unable to control the pest. Monitoring of test orchards showed
that after about 6\textendash{}18 years from the introduction of \emph{T.
sinensis}, the pest population declined to levels as low as to be
practically undetectable, giving rise to claims of success in biologically
controlling the infestation \citep{Moriya89,Murakami01,Moriya03}.
However, continuous monitoring of the first release site over 25 years
shows three successive peaks in the population of \emph{D. kuriphilus},
shortly followed by peaks in the population of \emph{T. sinensis}
(Moriya, personal communication). In the USA, several Asian \emph{Torymus}
species were released in 1977 in southeastern Georgia, but the release
was not followed by any monitoring. The first accounts of the successful
establishment of \emph{T. sinensis} in the United States were published
only thirty years later \citep{CooperRieske07,Rieske07}. In spite
of the abundant presence of \emph{T. sinensis}, and of \emph{Ormyrus
labotus} Walker (a native insect that was shown to easily parasitize
\emph{D. kuriphilus} galls), the pest could be found in most of the
southern Appalachian range, with satellite infestations in Ohio and
Pennsylvania. 

The European chestnut (\emph{Castanea sativa} Mill.) is one of the
most important broad-leaved species in Italy: chestnut stands amount
to 788,400 hectares, which represents 9\% of the Italian forests \citep{Graziosi08}.
Due to the report of the gall wasp in 2002 and in consideration of
the long-established economic importance of chestnut throughout the
country for fruit and wood production, a collaboration was started
with Japanese researchers and a biological control program was initiated
in 2005 with the release in infested orchards of Japan--imported \emph{T.
sinensis} specimens \citep{Quacchia08}. Following the Japanese and
Italian experiences, reporting the establishment of a sizable population
of \emph{T. sinensis} vigorously parasitizing the galls of \emph{D.
kuriphilus}, recent releasing programs were performed in Croatia,
France and Hungary \citep{Borowiec14,Mato=000161evi=00010714}, as
well as test releases in Spain and Portugal (Associação Portuguesa
da Castanha, personal communication).

Although in Europe there exist several native species of Hymenoptera
capable of parasitizing \emph{D. kuriphilus} galls, all of them have
a very large host range, and suffer by a mismatch between their emergence
times and the development of the galls. They are therefore unable
to act effectively as biological control agents \citep{Quacchia13,Alma14}.

In the present paper we develop a mathematical model of the interaction
between \emph{T. sinensis} and \emph{D. kuriphilus}. aiming at developing
a tool for understanding and evaluating the effectiveness of biological
control programs based on the release of \emph{T. sinensis} in woods
and orchards infested by \emph{D. kuriphilus}. 

In particular we would like to investigate whether \emph{T. sinensis}
should be expected to be able, alone, to cause the local extinction
of \emph{D. kuriphilus}, or maintain its population to levels as low
as to produce no harm, or if such expectations are over optimistic.
The fact that \emph{T. sinensis} is extremely well synchronized with
\emph{D. kuriphilus}, that outside China it acts almost perfectly
as host-specific, and that in Europe its abundance appears to be limited
only by the availability of its host, with a very low mortality during
all its life stages, allows hopes for a rapid, complete, and permanent
control of the pest. However, the experiences of both Japan and the
USA warn that the effectiveness of \emph{T. sinensis} might be less
perfect than one would wish it to be. In the case of Japan the imperfect
control of \emph{D. kuriphilus} has been ascribed to a high mortality
of \emph{T. sinensis} by hyperparasitism \citep{Murakami91}. In Europe
hyperparasitism is only occasional \citep{Quacchia13}, which leaves
more room for hopes of obtaining a control.

In order to have a flexible tool, our model, in its full form, is
a spatially-explicit one, which also explicitly describes, both for
\emph{T. sinensis} and for \emph{D. kuriphilus,} the seasonal time
evolution of the adult insect population, and the competition for
finding the egg deposition sites. In a spatially--homogeneous situation
the model may be rigorously reduced to an iterated map quantifying
the egg density of the two species, whose properties are studied with
a combination of analytic and numerical techniques. The full, spatially--explicit
model is studied by means of numerical simulations in one and two
spatial dimensions. The comparison between the dynamics of the iterated
maps and of the full model suggests a diffusion-based mechanisms that
may give rise, under certain conditions, to repeated waves of full
infestation followed by near disappearance of the pest and of its
parasitoid, on time scales that depend non only on the physiological
and ecological parameters, but also on the size and geometry of the
wood.

The rest of the paper is organized as follows: the mathematical model
is developed in section \ref{sec:The-model}; the results obtained
from the model are reported in detail in section \ref{sec:Results};
finally they are summarized in section \ref{sec:Discussion-and-conclusions},
together with some speculative considerations. Section \ref{sec:Appendix}
is an appendix containing mathematical analyses in support of statements
appearing in sections \ref{sec:The-model} and \ref{sec:Discussion-and-conclusions}.

\section{The model\label{sec:The-model}}

\subsection{\label{sub:Equations-for-gall-wasp}Equations for the gall wasp}

We describe the population of adult gall wasps during the summer of
the year $n$ as a field $U_{n}$ quantifying the density (that is,
number of insects per unit area) of egg--carrying \emph{D. kuriphilus}
adults. By ``density of egg--carrying adults'' we mean that an adult
that has not yet laid any eggs contributes by a whole unit in the
computation of this density, an adult that has laid, say, half its
eggs contributes by half a unit, and one that has laid all its eggs
does not contribute at all, even if it is still alive. Thus, calling
$N_{D}$ the maximum number of eggs that can be laid by a typical
\emph{D. kuriphilus} adult under optimal conditions, then $N_{D}U_{n}(\boldsymbol{x},t)$
is the number of eggs per unit area present at the location $\boldsymbol{x}$
and time $t$ that can still potentially be laid. 

We shall also need a second field, $V_{n}$, that quantifies the density
of eggs laid in chestnut buds. Because \emph{D. kuriphilus} may only
lay eggs on chestnuts buds, and at most $M$ eggs per bud, then the
density of laid eggs in any location $\boldsymbol{x}$ is always at
most $M\beta_{n}(\boldsymbol{x})$, where $\beta_{n}$ is the density
of chestnut buds on the $n-$th year. In any case, the maximum density
of laid eggs cannot exceed the quantity 
\begin{equation}
V_{max}=M\beta_{max}\label{eq:Vmax}
\end{equation}
 where the constant $\beta_{max}$ is the maximum density of buds
attainable in a chestnut wood under optimal conditions.

At the beginning of each season, the density of both the gall wasps
and of their laid eggs are zero:

\begin{equation}
\begin{cases}
U_{n}(\boldsymbol{x},0) & =0\\
V_{n}(\boldsymbol{x},0) & =0
\end{cases}.\label{eq:UV_inicond}
\end{equation}
As the season progresses, from the galls formed during the previous
season, the wasps gradually emerge. For simplicity we shall assume
a constant emergence rate:
\begin{equation}
\mathrm{emergence\, rate}=\frac{\eta V_{n-1}(\boldsymbol{x},T_{D})}{T_{D}}\label{eq:emergence_rate}
\end{equation}
where $T_{D}$ is the length of the egg deposition season, and the
non-dimensional parameter $\eta\in(0,1]$ is the survival rate during
the overwintering. More precisely, $\eta V_{n-1}(\boldsymbol{x},T_{D})\, dA$
is the number of \emph{D. kuriphilus} adults that emerge during the
$n-$th season from an area $dA$ centered around the location $\boldsymbol{x}$.
Taking into account that chestnut gall wasps reproduce by thelytokous
parthenogenesis \citep{Murakami81}, and have a low natural mortality
of eggs and larvae, we expect the numerical value of $\eta$ to be
close to one. More in detail, the primary mortality factors for \emph{D.
kuriphilus} are parasitism, gall-chamber invading fungi and failure
of adult gall wasp to emerge \citep{CooperRieske10}, but from our
experience all these processes have effects so mild to be almost negligible
(authors' personal observation).

Individual gall wasps do not survive for more than a few days. Therefore
we need to introduce a sink term representing their mortality rate.
We are not aware of any evidence in the literature of important exogenous
factors affecting the mortality of adult gall wasps. Thus, taking
individual deaths as independent from each other, the rate of deaths
per unit area is likely to be proportional to the density of the population,
suggesting the following simple choice for the death rate term
\begin{equation}
\mathrm{death\, rate}=-\frac{U_{n}(\boldsymbol{x},t)}{a}\label{eq:death_rate}
\end{equation}
where $a$ is the typical adult life span (up to ten days: \citealp{EFSA10}).

We shall assume that during the egg-laying season the gall wasps move
randomly, diffusing isotropically in the forest. Although there is
evidence of a response of \emph{D. kuriphilus} to olfactory cues in
the choice of a host twig, this was observed spatial scales shorter
than a meter \citep{Germinara11}. On much larger scales there is
no evidence of anisotropic motion of the gall wasps, nor it should
be expected. Following olfactory cues in a turbulent environment,
such as a wood canopy, is a very challenging task when there is a
single odor source \citep{Balkovsky&Shraiman}. In the presence of
multiple sources it is very unlikely that an insect can consistently
and reliably exploit olfactory cues on long range. For example, in
the case of the parasitoid wasp \emph{Diachasmimorpha juglandis} Muesebeck\emph{,}
it was verified that it preferred to use visual cues rather than olfactory
ones for locating the walnut fruit husks where its host may be found
\citep{Henneman02}. In the case of \emph{D. kuriphilus}, the available
visual cues are also short-range: chestnut buds are not visible from
more than a few meters away. Therefore, we consider reasonable to
assume that the large-scale motion (that is, on distances larger than
the size of individual trees) of \emph{D. kuriphilus} adults is aimless
and random, and thus it should be modeled by a Laplacian diffusion
operator (we shall further discuss this issue in section \ref{sec:Discussion-and-conclusions}).

When the egg-carrying adults find available buds (that is buds that
are not already fully saturated by other eggs), they quickly lay one
or more eggs, thus reducing the number of available deposition sites.
The rate of egg deposition of an individual will be proportional to
the density of available eggs deposition sites, which, in the model,
is expressed as $M\beta_{n}(\boldsymbol{x})-V_{n}(\boldsymbol{x},t)$.
It would be more accurate to assume that the egg deposition rate is
a Holling's type II function of the available egg deposition sites.
However, our observations suggest that, for \emph{D. kuriphilus,}
the handling time (the time actually spent laying eggs) is just a
tiny fraction of the search time (which is comparable with the adult
life span). When the handling time is negligible, the Holling's type
II function tends to a simple proportionality between the deposition
rate and the density of available deposition places \citep[see e.g.][p.163]{Vandermeer&Goldberg13}.
Accordingly, the egg deposition rate of the whole population is taken
as proportional to the product of the density of available sites by
the density of the adult population, as in the following expression
\begin{equation}
\mathrm{egg\, deposition\, rate}={\displaystyle r_{D}\frac{M\beta_{n}(\boldsymbol{x})-V_{n}(\boldsymbol{x},t)}{V_{max}}U_{n}(\boldsymbol{x},t)}.\label{eq:egg_deposition_rate}
\end{equation}
It is possible to give a reasonable estimate for the proportionality
constant $r_{D}$ that appears in the expression above. In fact, we
must assume that in optimal conditions (that is, if $V_{n}=0$, $\beta_{n}(\boldsymbol{x})=\beta_{max}$
and thus the deposition rate reduces to $r_{D}U_{n}$) every adult
gall wasp must be able to lay all its $N_{D}$ eggs in a time interval
roughly equal to its adult life span $a$. This would imply that 
\begin{equation}
r_{D}=\frac{N_{D}}{a}\label{eq:r_value}
\end{equation}

By adding together all the processes discussed in this section we
arrive to the following model that describes the time evolution of
the $U_{n}$ and $V_{n}$ fields during the $n-$th season. 
\begin{equation}
\begin{cases}
{\displaystyle \frac{\partial}{\partial t}U_{n}(\boldsymbol{x},t)\,=} & {\displaystyle D_{D}\nabla^{2}U_{n}(\boldsymbol{x},t)-\frac{1}{a}\frac{M\beta_{n}(\boldsymbol{x})-V_{n}(\boldsymbol{x},t)}{V_{max}}U_{n}(\boldsymbol{x},t)}\\
 & \vphantom{\frac{\Big\{}{\Big\{}}-\frac{1}{a}U_{n}(\boldsymbol{x},t)+{\displaystyle \frac{\eta V_{n-1}(\boldsymbol{x},T_{D})}{T_{D}}}\\
{\displaystyle \frac{\partial}{\partial t}V_{n}(\boldsymbol{x},t)\,=} & \vphantom{\frac{\Big\{}{\Big\{}}{\displaystyle \frac{N_{D}}{a}\frac{M\beta_{n}(\boldsymbol{x})-V_{n}(\boldsymbol{x},t)}{V_{max}}U_{n}(\boldsymbol{x},t)}
\end{cases}\label{eq:the_cinipide_model}
\end{equation}
where $D_{D}$ is the diffusivity of the gall wasps, and all other
symbols have already been defined. Note that the egg deposition rate,
that appears as the only term in the right-hand side of the equation
for $V_{n}$, also appears in the equation for $U_{n}$ with a minus
sign and divided by $N_{D}$. This is because, as discussed above,
the contribution of each individual to the density $U_{n}$ is weighted
by the fraction of eggs that it carries. 

The problem (\ref{eq:the_cinipide_model}) with the initial conditions
(\ref{eq:UV_inicond}) is not closed, because no rule was specified
for the time evolution of the bud density $\beta_{n}$. In the presence
of a developed infestation the health of the chestnut trees progressively
deteriorates, and the bud density may decrease. This is a slow process,
whose details are largely unknown \citep{Kato97}. If the model were
used to perform detailed, realistic year--by--year forecasts of the
spreading of \emph{D. kuriphilus}, the best results would be obtained
by measuring the density $\beta_{n}$ by means of direct surveys of
the orchards and coppices under study. In this paper, in order to
assess and understand the main features of the solutions of the model's
equations, we shall use the strong simplifying assumption that the
density of buds is always constant, and equal to $\beta_{max}$. 

It is convenient to make non--dimensional the dependent variables,
by defining $u_{n}=U_{n}/\left(\eta V_{max}\right)$ and $v_{n}=V_{n}/V_{max}$.
Note that $v_{n}\in[0,1]$ and that $M\beta_{n}/V_{max}=1$, because
of the simplifying assumption $\beta_{n}=\beta_{max}$. Likewise,
it is convenient to use non--dimensional variables also for time and
space. These are defined as: $\tilde{t}=t/T_{D}$ and $\tilde{\boldsymbol{x}}=\boldsymbol{x}/\sqrt{D_{D}T_{D}}$.
Thus the equations (\ref{eq:the_cinipide_model}) become (for typographical
brevity in the following we shall omit the tildes on the non--dimensional
independent variables) 
\begin{equation}
\begin{cases}
{\displaystyle \frac{\partial}{\partial t}u_{n}}(\boldsymbol{x},t) & ={\displaystyle \nabla^{2}u_{n}(\boldsymbol{x},t)-\mu\left(2-v_{n}(\boldsymbol{x},t)\right)u_{n}(\boldsymbol{x},t)}+v_{n-1}(\boldsymbol{x},1)\\
{\displaystyle \frac{\partial}{\partial t}v_{n}(\boldsymbol{x},t)} & ={\displaystyle E_{D}\mu\left(1-v_{n}(\boldsymbol{x},t)\right)u_{n}(\boldsymbol{x},t)}
\end{cases}\label{eq:cinipide_model_uv_rescaled}
\end{equation}
where $\mu=T_{D}/a$, $E_{D}=\eta N_{D}$ and $t\in[0,1]$. For each
$n$, the equations (\ref{eq:cinipide_model_uv_rescaled}) are subject
to the conditions 
\begin{equation}
\begin{cases}
u_{n}(\boldsymbol{x},0) & =0\\
v_{n}(\boldsymbol{x},0) & =0
\end{cases}\label{eq:Cinipede_initial_conditions}
\end{equation}
This is a piecewise smooth initial value problem, characterized by
two free parameters: $E_{D}$ and $\mu$. The first one is the maximum
number of eggs that can be laid by a \emph{D. kuriphilus} adult, multiplied
by the overwintering mortality (which does not appear elsewhere in
the non--dimensional equations); the second is the reciprocal of the
adult life span, measured in the non--dimensional time units. A further
important parameter is the size, in non--dimensional units, of the
domain $\Omega$, that is the chestnut--covered area on which $U_{n}$
and $V_{n}$ are defined. The equations (\ref{eq:cinipide_model_uv_rescaled})
and the conditions (\ref{eq:Cinipede_initial_conditions}) must be
complemented by suitable boundary conditions describing the behavior
of the gall wasps when they find themselves at the edge of the wood.
We are not aware of any published work on this issue. It is very likely
that a small fraction of the gall wasps would venture outside a chestnut
orchard or coppice, spilling over adjacent regions. For simplicity,
here we assume that any gall--wasp that were to leave the domain $\Omega$
would promptly change its course, returning inside the chestnut--populated
area. In this idealized situation there would be no flux of wasps
across the edges of $\Omega$, and therefore the appropriate boundary
conditions for $U_{n}$ would be
\begin{equation}
\left.\hat{\boldsymbol{n}}\cdot\nabla u_{n}\right|_{\partial\Omega}=0\label{eq:BC_no_U_flux}
\end{equation}
where $\partial\Omega$ denotes the line delimiting the boundary of
$\Omega$, and $\hat{\boldsymbol{n}}$ represents the outward unit
vector perpendicular to $\partial\Omega$. 

The no--flux boundary condition (\ref{eq:BC_no_U_flux}) is of particular
interest because it allows for homogeneous solutions, that is, solutions
in which the densities $u_{n}$ and $v_{n}$ are constant in space
(but not in time). In particular, it is straightforward to verify
that if $s_{0}$ is taken as a constant, then at all later times $t$
and seasons $n$, $u_{n}$ and $v_{n}$ do not depend on $\boldsymbol{x}$,
and the problem (\ref{eq:cinipide_model_uv_rescaled},\ref{eq:Cinipede_initial_conditions},\ref{eq:BC_no_U_flux})
reduces to the following chain of ordinary differential equations:
\begin{equation}
\begin{cases}
\dot{u}_{n}(t) & =v_{n-1}(1)-\mu\left(2-v_{n}(t)\right)u_{n}(t)\\
\dot{v}_{n}(t) & =E_{D}\mu\left(1-v_{n}(t)\right)u_{n}(t)\\
u_{n}(0) & =0\\
v_{n}(0) & =0
\end{cases}\label{eq:ODEs}
\end{equation}
where the dot denotes differentiation with respect to time. The solution
of these nonlinear equations cannot be expressed in terms of simple
functions. However, a formal calculation shows that the year--over--year
dynamics of the egg density can be well approximated by the following
simple map (see Appendix \ref{sub:Bounds} for details)
\begin{figure}
\begin{centering}
\includegraphics[width=0.85\textwidth]{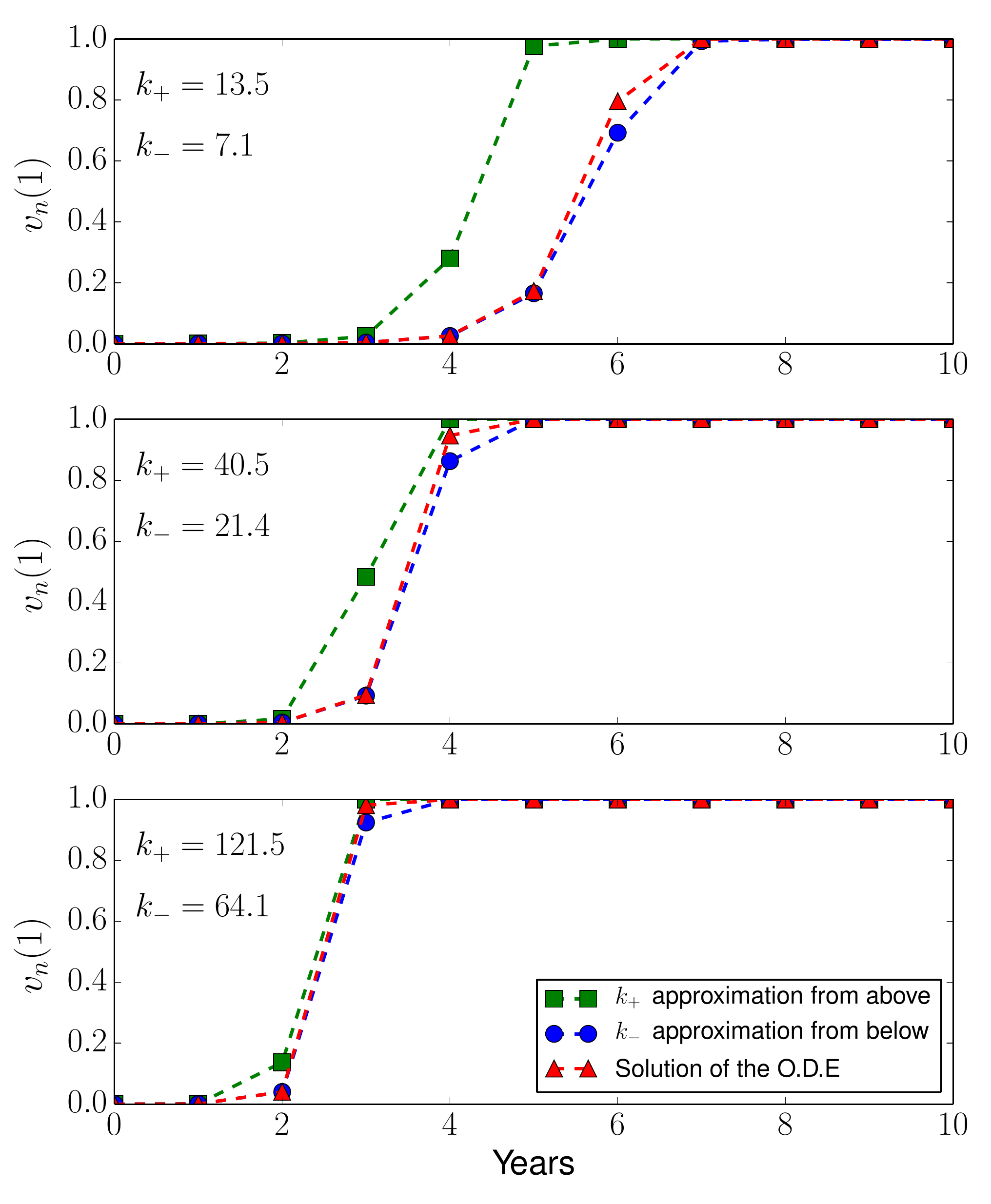}
\par\end{centering}

\caption{\label{fig:Maps_vs_ODE}Comparison of the end--of--season egg densities
given by Skellam's map (\ref{eq:Skellam-model}) and by a numerical
solution of the spatially--homogeneous model (\ref{eq:ODEs}) for
different values of the parameter $k_{\pm}$ defined in (\ref{eq:Skellam_constant}).
In all the computations we have used $N_{D}=150$, $\mu=10$, $v_{0}(1)=10^{-5}$.
The overwintering survival rates are, from top to bottom, $\eta=0.1,\,0.3,\,0.9$.
The first two values are unrealistically low and are meant just to
illustrate the properties of the two approximations. The last value
is considered to be realistic in the European setting.}
\end{figure}
 
\begin{equation}
v_{n}=1-e^{-kv_{n-1}}\label{eq:Skellam-model}
\end{equation}
where the egg densities $v_{n}$, $v_{n-1}$ are evaluated at time
$t=1$ (corresponding to the end of the $n-$th and $(n-1)-$th seasons).
The positive constant $k$ that appears in the exponential is given
by one of the following two expressions
\begin{equation}
\begin{cases}
k_{+}=\frac{E_{D}}{\mu}\left(e^{-\mu}+\mu-1\right) & \mathrm{approx.\, from\, above,}\\
k_{-}=\frac{E_{D}}{4\mu}\left(e^{-2\mu}+2\mu-1\right) & \mathrm{approx.\, from\, below}.
\end{cases}\label{eq:Skellam_constant}
\end{equation}
On choosing $k=k_{-}$ we obtain an approximation from below, which
is highly accurate when the previous year egg density $v_{n-1}$ is
appreciably smaller than 1. Choosing $k=k_{+}$ one has an approximation
from above, that captures the dynamics more accurately when the egg
density $v_{n-1}$ is close to 1. (see Appendix \ref{sub:Bounds}
for details, and Figure \ref{fig:Maps_vs_ODE}).

We observe that the iterated map (\ref{eq:Skellam-model}) is the
well-known model of Skellam \citep{Skellam51,Brannstrom05}, that
describes the population dynamics of univoltine insects in a regime
of contest competition.

\subsection{\label{sub:Equations-for-Torymus}Equations for \emph{T. sinensis}}

\emph{Torymus} adults emerge from vacated galls in spring. There appears
to be a good degree of synchronism in the emergence process, so that
the great majority of all the individuals appear in a time span of
a few days (authors' personal observation). After mating, the egg-carrying
females look for intact galls into which they lay (usually) one egg
per chamber \citep{Piao92}. Each female initially carries about $N_{T}\approx70$
eggs. In outdoor conditions the adult lifetime of \emph{T. sinensis}
is at least 37 days \citep{Piao92}. For modeling purposes we shall
take the emergence as an instantaneous process, and attribute the
same life--span to all the individuals, so that they all die together. 

In the following we will denote with $P_{n}$ the density of the egg-carrying
\emph{T. sinensis} females and with $Q_{n}$ the density of the eggs
already laid, during the $n-$th season. Just as in our model of \emph{D.
kuriphilus,} we shall use the following expression for the egg deposition
rate of \emph{T. sinensis} 
\begin{equation}
\mathrm{egg\, deposition\, rate}={\displaystyle r_{T}\frac{\eta V_{n-1}(\boldsymbol{x},T_{D})-Q_{n}(\boldsymbol{x},t)}{V_{max}}P_{n}(\boldsymbol{x},t)}\label{eq:Torymus_egg_deposition_rate}
\end{equation}
Also in this case, in principle, the rate should be expressed through
a Holling's type II functional response. But the oviposition time
of \emph{T. sinensis} is very short (a few minutes, authors' personal
observation) in comparison with its search time. Thus, as we argued
in the case of the gall wasp, the deposition rate must be proportional
to the product of the density of egg-carrying \emph{T. sinensis} females
with the density of the sites where oviposition is possible. The latter
is given by the difference between the density of gall wasp eggs laid
during the previous season and turned into larvae (namely $\eta V_{n-1}$)
and the density of \emph{T. sinensis} eggs already laid. Here $r_{T}/V_{max}$
is the proportionality constant. As for the gall--wasp we should assume
that every female \emph{Torymus}, in optimal conditions (that is,
$\eta V_{n-1}=V_{max}$ and $Q_{n}=0$), should be able to lay all
its $N_{T}$ eggs during its life span $T_{T}$. Thus, we assume
\begin{equation}
r_{T}=\frac{N_{T}}{T_{T}}.\label{eq:Torymus_deposition_rate}
\end{equation}

Also for \emph{T. sinensis} it is know that it responds to olfactory
and visual cues at short ranges \citep{Graziosi13}. On longer distances,
the same considerations already mentioned for the gall--wasp apply:
the overall motion of an adult \emph{T. sinensis} during its life
span should be random and aimless, and therefore a Laplacian diffusion
process, characterized by a constant diffusivity $D_{T}$, should
be the appropriate model. 

Therefore we may describe the dynamics of a population of \emph{T.
sinensis} during the $n-$th season with the following equations:

\begin{equation}
\begin{cases}
{\displaystyle \frac{\partial}{\partial t}P_{n}(\boldsymbol{x},t)} & ={\displaystyle D_{T}\nabla^{2}P_{n}(\boldsymbol{x},t)-\frac{1}{T_{T}}\frac{\eta V_{n-1}(\boldsymbol{x},T_{D})-Q_{n}(\boldsymbol{x},t)}{V_{max}}P_{n}(\boldsymbol{x},t)}\\
{\displaystyle \frac{\partial}{\partial t}Q_{n}(\boldsymbol{x},t)} & ={\displaystyle \frac{N_{T}}{T_{T}}\frac{\eta V_{n-1}(\boldsymbol{x},T_{D})-Q_{n}(\boldsymbol{x},t)}{V_{max}}P_{n}(\boldsymbol{x},t)}
\end{cases}.\label{eq:Torymus_equations}
\end{equation}
Here the time $t=0$ corresponds to the simultaneous emergence of
the adult \emph{Torymus}. The equations are valid up to $t=T_{T}$,
corresponding to the end of the \emph{Torymus} season, when all the
adults die. Just as we did for the gall wasp, the rate of change of
the laid egg density $Q_{n}$ is equal to the egg deposition rate.
This rate is also divided by $N_{T}$ and subtracted from the equation
for the rate of change of the density $P_{n}$ of the \emph{T. sinensis}
females, because the density of adult females is weighted by the number
of eggs that each adult carries.

The equations (\ref{eq:Torymus_equations}) are subject to the initial
conditions 
\begin{equation}
\begin{cases}
P_{n}(\boldsymbol{x},0) & =\gamma Q_{n-1}(\boldsymbol{x},T_{T})\\
Q_{n}(\boldsymbol{x},0) & =0
\end{cases}\label{eq:Torymus_initial_conditions}
\end{equation}
 The initial density of \emph{T. sinensis} females is not zero because
we assumed the instantaneous emergence of all the adults. The constant
$\gamma$ accounts for the sex ratio of \emph{T. sinensis}, and for
the mortality rate of the overwintering larvae. Male and female have
roughly the same probability to emerge from a fertilized egg of \emph{T.
sinensis} \citep{Ferracini15b} and the overwintering mortality is
believed to be very low (author's personal observation), thus we shall
use values of $\gamma$ smaller than, but close to $1/2$. \emph{T.
sinensis} females that are not able to mate may still lay their unfertilized
eggs, from which will emerge males, by arrhenotokous parthenogenesis.
Therefore, if the density of \emph{T. sinensis} drops to very low
levels, in the next season the sex ratio will be skewed in favor of
the males, resulting in an improved mating probability for the remaining
females. In its present form, our model does not include this mechanism.
However, we also do not model explicitly the mating process: all the
females are implicitly considered to be fertilized at the moment of
their emergence. Thus we are already overestimating the mating probability
of the females, and we feel that, at this stage, further complications
may be unnecessary. For the same reason, \emph{T. sinensis} is modeled
as a strictly univoltine species. The recent observations of an extended
diapause of a few \emph{Torymus} individuals, in a controlled setting
\citep{Ferracini15b}, does not yet allow a quantitative assessment
of the importance (if any) of this process for the dynamics of the
population in the wild. Thus we postpone the inclusion of these processes
for a possible future improved version of the model.

It is convenient to rewrite the model by using the non--dimensional
densities $p_{n}=P_{n}/(\gamma\eta V_{max})$, $q_{n}=Q_{n}/(\eta V_{max})$,
and the same non--dimensional space and time variables already used
for the gall--wasp equations. The equations (\ref{eq:Torymus_equations})
then become 
\begin{equation}
\begin{cases}
{\displaystyle \frac{\partial}{\partial t}p_{n}(\boldsymbol{x},t)\,=} & {\displaystyle \delta\nabla^{2}p_{n}(\boldsymbol{x},t)-\tau^{-1}\left(v_{n-1}(\boldsymbol{x},1)-q_{n}(\boldsymbol{x},t)\right)p_{n}(\boldsymbol{x},t)},\\
{\displaystyle \frac{\partial}{\partial t}q_{n}(\boldsymbol{x},t)\,=} & E_{T}\tau^{-1}\left(v_{n-1}(\boldsymbol{x},1)-q_{n}(\boldsymbol{x},t)\right)p_{n}(\boldsymbol{x},t),
\end{cases}\label{eq:Torymus_equations_nondim}
\end{equation}
where we have defined the diffusivity ratio $\delta=D_{T}/D_{D}$,
the non--dimensional \emph{T. sinensis} season length $\tau=T_{T}/(\eta T_{D})$,
and the effective egg number $E_{T}=\gamma N_{T}$. The initial conditions
(\ref{eq:Torymus_initial_conditions}) become
\begin{equation}
\begin{cases}
p_{n}(\boldsymbol{x},0)\,= & q_{n-1}(\boldsymbol{x},\eta\tau),\\
q_{n}(\boldsymbol{x},0)\,= & 0.
\end{cases}\label{eq:Torymus_initial_conditions-nondim}
\end{equation}
By imposing no--flux boundary conditions on $p_{n}$, and looking
for homogeneous solutions, the equations (\ref{eq:Torymus_equations_nondim})
together with the initial conditions, yield a set of ordinary differential
equations whose solution is given in Appendix (\ref{sub:Exact_solutions_Torymus}).
By evaluating the solution at the time corresponding to the end of
the \emph{Torymus} season, that is at the non--dimensional time $t=\eta\tau$,
we obtain the following map: 

\begin{equation}
q_{n+1}=\begin{cases}
{\displaystyle \frac{E_{T}v_{n}q_{n}\left(1-e^{\eta\left(E_{T}q_{n}-v_{n}\right)}\right)}{v_{n}-E_{T}q_{n}e^{\eta\left(E_{T}q_{n}-v_{n}\right)}}}, & E_{T}q_{n}\neq v_{n}\\
\frac{v_{n}^{2}}{v_{n}+\eta^{-1}}, & E_{T}q_{n}=v_{n}
\end{cases}\label{eq:Torymus_map}
\end{equation}
where the egg densities $q_{n+1}$, $q_{n}$ and $v_{n}$ are evaluated
at the end of their respective seasons. Albeit complicated--looking,
the right--hand side of the map is a smooth function of its parameters,
even for $E_{T}q_{n}=v_{n}$. In particular, it is a growing function
of $q_{n}$, and, for realistic values of $E_{T}$ and $\eta$, it
rapidly approaches the horizontal asymptote $q_{n+1}\to v_{n}$. Therefore,
the map (\ref{eq:Torymus_map}), and thus the underlying equations
(\ref{eq:Torymus_equations}), are a model that describes a contest
competition process among the individuals of \emph{T. sinensis} \citep{Brannstrom05}.

\subsection{The complete model}

The equations for \emph{T. sinensis,} discussed in the previous subsection,
already depend on the density of \emph{D. kuriphilus} eggs laid in
the previous year. In order to have a fully coupled model, we only
need to incorporate the parasitism of \emph{T. sinensis} in the equations
for \emph{D. kuriphilus} discussed in sec \ref{sub:Equations-for-gall-wasp}.
This is easily accomplished by observing that parasitized larvae of
\emph{D. kuriphilus} simply won't give rise to adults. Therefore we
need to change the emergence rate (\ref{eq:emergence_rate}) with
\begin{equation}
\mathrm{emergence\, rate}=\frac{\eta V_{n-1}(\boldsymbol{x},T_{D})-Q_{n}(\boldsymbol{x},T_{T})}{T_{D}}.\label{eq:emergence_rate-1}
\end{equation}
The complete model, using the non--dimensional variables, then reads
\begin{equation}
\begin{cases}
{\displaystyle \frac{\partial}{\partial t}p_{n}(\boldsymbol{x},t)} & ={\displaystyle \delta\nabla^{2}p_{n}(\boldsymbol{x},t)-\tau^{-1}\left(v_{n-1}(\boldsymbol{x},1)-q_{n}(\boldsymbol{x},t)\right)p_{n}(\boldsymbol{x},t)}\\
{\displaystyle \frac{\partial}{\partial t}q_{n}(\boldsymbol{x},t)} & =E_{T}\tau^{-1}\left(v_{n-1}(\boldsymbol{x},1)-q_{n}(\boldsymbol{x},t)\right)p_{n}(\boldsymbol{x},t)\\
{\displaystyle \frac{\partial}{\partial t}u_{n}}(\boldsymbol{x},t) & ={\displaystyle \nabla^{2}u_{n}(\boldsymbol{x},t)-\mu\left(2-v_{n}(\boldsymbol{x},t)\right)u_{n}(\boldsymbol{x},t)}+v_{n-1}(\boldsymbol{x},1)-q_{n}(\boldsymbol{x},\eta\tau)\\
{\displaystyle \frac{\partial}{\partial t}v_{n}(\boldsymbol{x},t)} & ={\displaystyle E_{D}\mu\left(1-v_{n}(\boldsymbol{x},t)\right)u_{n}(\boldsymbol{x},t)}\\
p_{n}(\boldsymbol{x},0) & =q_{n-1}(\boldsymbol{x},\eta\tau)\\
q_{n}(\boldsymbol{x},0) & =0\\
u_{n}(\boldsymbol{x},0) & =0\\
v_{n}(\boldsymbol{x},0) & =0.
\end{cases}\label{eq:complete_model}
\end{equation}
In the case of space--independent solutions, the dynamic of this model
is well approximated by the following map
\begin{equation}
\begin{cases}
q_{n+1} & =\begin{cases}
{\displaystyle \frac{E_{T}v_{n}q_{n}\left(1-e^{\eta\left(E_{T}q_{n}-v_{n}\right)}\right)}{v_{n}-E_{T}q_{n}e^{\eta\left(E_{T}q_{n}-v_{n}\right)}}}, & E_{T}q_{n}\neq v_{n}\\
\frac{v_{n}^{2}}{v_{n}+\eta^{-1}}, & E_{T}q_{n}=v_{n}
\end{cases}\\
v_{n+1} & =1-e^{-k\left(v_{n}-q_{n+1}\right)}
\end{cases}.\label{eq:complete_map}
\end{equation}
that describes the year--over--year change of the end--of--season
density of \emph{T. sinensis} and \emph{D. kuriphilus} eggs.

\begin{table}
\noindent \begin{centering}
\begin{tabular}{|c|c|c|c|}
\hline 
$D_{D}$ & $0.77\,\mathrm{km^{2}d^{-1}}$ & %
\begin{minipage}[t]{0.3\textwidth}%
Diffusion coefficient of \emph{}\\
\emph{D. kuriphilus.}%
\end{minipage} & %
\begin{minipage}[t]{0.4\textwidth}%
See sec. \ref{sub:Space-dependent-dynamics}.%
\end{minipage}\tabularnewline
\hline 
$a$ & %
\begin{minipage}[t]{0.2\textwidth}%
2--10 days

2--3~ days%
\end{minipage} & %
\begin{minipage}[t]{0.3\textwidth}%
Adult life span of \\
\emph{D. kuriphilus.}%
\end{minipage} & %
\begin{minipage}[t]{0.4\textwidth}%
\citet{EFSA10};

\citet{Graziosi14}.%
\end{minipage}\tabularnewline
\hline 
$M$ & 20--30 eggs bud$^{-1}$ & %
\begin{minipage}[t]{0.3\textwidth}%
number of eggs of \emph{D. kuriphilus} that can be laid on a bud.%
\end{minipage} & %
\begin{minipage}[t]{0.4\textwidth}%
\citet{EFSA10}.%
\end{minipage}\tabularnewline
\hline 
$\beta_{max}$ & $2\cdot10^{6}$buds ha$^{-1}$ & %
\begin{minipage}[t]{0.3\textwidth}%
Maximum density of chestnut buds.%
\end{minipage} & %
\begin{minipage}[t]{0.4\textwidth}%
\citet{Bounous14}.%
\end{minipage}\tabularnewline
\hline 
$\eta$ & 0.5--0.98 & %
\begin{minipage}[t]{0.3\textwidth}%
Fraction of \emph{D. kuriphilus} larvae surviving after overwintering.%
\end{minipage} & %
\begin{minipage}[t]{0.4\textwidth}%
\citet{CooperRieske07};

\citet{Quacchia13}.%
\end{minipage}\tabularnewline
\hline 
$T_{D}$ & 30--50 days & %
\begin{minipage}[t]{0.3\textwidth}%
Length of the egg deposition season for \emph{D. kuriphilus.}%
\end{minipage} & %
\begin{minipage}[t]{0.4\textwidth}%
\citet{Eppo05}.%
\end{minipage}\tabularnewline
\hline 
$N_{D}$ & 100--300 & %
\begin{minipage}[t]{0.3\textwidth}%
Number of eggs per adult of \emph{D. kuriphilus.}%
\end{minipage} & %
\begin{minipage}[t]{0.4\textwidth}%
\citet{Graziosi14}.%
\end{minipage}\tabularnewline
\hline 
$D_{T}$ & unknown & %
\begin{minipage}[t]{0.3\textwidth}%
Diffusion coefficient of \emph{}\\
\emph{T. sinensis}.%
\end{minipage} & %
\begin{minipage}[t]{0.4\textwidth}%
See sec. \ref{sub:Space-dependent-dynamics-of-complete-model} %
\end{minipage}\tabularnewline
\hline 
$T_{T}$ & 37 days or more & %
\begin{minipage}[t]{0.3\textwidth}%
Length of the egg deposition season for \emph{T. sinensis.}%
\end{minipage} & %
\begin{minipage}[t]{0.4\textwidth}%
\citet{Piao92}.%
\end{minipage}\tabularnewline
\hline 
$N_{T}$ & 71 & %
\begin{minipage}[t]{0.3\textwidth}%
Number of eggs per adult female of \emph{T. sinensis}.%
\end{minipage} & %
\begin{minipage}[t]{0.4\textwidth}%
\citet{Piao92}.%
\end{minipage}\tabularnewline
\hline 
$\gamma$ & 0.25--0.45 & %
\begin{minipage}[t]{0.3\textwidth}%
Fraction of \emph{T. sinensis} larvae that are female and survive
after overwintering.%
\end{minipage} & %
\begin{minipage}[t]{0.4\textwidth}%
\citet{Piao92}

Author's unpublished observations%
\end{minipage}\tabularnewline
\hline 
\end{tabular}
\par\end{centering}

\caption{\label{tab:Parameters}Parameters of the model and their likely value
or value range.}

\end{table}

\begin{table}
\noindent \begin{centering}
\begin{tabular}{|c|c|}
\hline 
\begin{minipage}[t]{0.35\textwidth}%
\[
u_{n}=\frac{U_{n}}{\eta M\beta_{max}}
\]
\end{minipage} & %
\begin{minipage}[t]{0.4\textwidth}%
Non--dimensional density of \emph{D. kuriphilus} adults during the
season $n$.%
\end{minipage}\tabularnewline
\hline 
\begin{minipage}[t]{0.35\textwidth}%
\[
v_{n}=\frac{V_{n}}{M\beta_{max}}
\]
\end{minipage} & %
\begin{minipage}[t]{0.4\textwidth}%
Non--dimensional density of \emph{D. kuriphilus} eggs laid during
the season $n$.%
\end{minipage}\tabularnewline
\hline 
\begin{minipage}[t]{0.35\textwidth}%
\[
p_{n}=\frac{P_{n}}{\gamma\eta M\beta_{max}}
\]
\end{minipage} & %
\begin{minipage}[t]{0.4\textwidth}%
Non--dimensional density of \emph{T. sinensis} adult females during
the season $n$.%
\end{minipage}\tabularnewline
\hline 
\begin{minipage}[t]{0.35\textwidth}%
\[
q_{n}=\frac{Q_{n}}{\eta M\beta_{max}}
\]
\end{minipage} & %
\begin{minipage}[t]{0.4\textwidth}%
Non--dimensional density of \emph{T. sinensis} eggs laid during the
season $n$.%
\end{minipage}\tabularnewline
\hline 
\begin{minipage}[b][1\totalheight][t]{0.35\textwidth}%
\[
\mu=\frac{T_{D}}{a}
\]
\end{minipage} & %
\begin{minipage}[t]{0.4\textwidth}%
Non--dimensional length of the egg deposition season of \emph{D. kuriphilus}.%
\end{minipage} \tabularnewline
\hline 
\begin{minipage}[t]{0.35\textwidth}%
\[
E_{D}=\eta N_{D}
\]
\end{minipage} & %
\begin{minipage}[t]{0.4\textwidth}%
Number of larvae per adult of \emph{D. kuriphilus} that survive the
overwintering in optimal conditions.%
\end{minipage}\tabularnewline
\hline 
\begin{minipage}[b][1\totalheight][t]{0.45\textwidth}%
\[
k=\begin{cases}
k_{+}=\frac{E_{D}}{\mu}\left(e^{-\mu}+\mu-1\right)\\
k_{-}=\frac{E_{D}}{4\mu}\left(e^{-2\mu}+2\mu-1\right)
\end{cases}
\]
\end{minipage} & %
\begin{minipage}[c][1\totalheight][t]{0.4\textwidth}%
Effective growth rate in the Skellam maps approximating from above
($k=k_{+}$) or from below ($k=k_{-}$) the year--over--year dynamics
of \emph{D. kuriphilus'} egg density.%
\end{minipage}\tabularnewline
\hline 
\begin{minipage}[t]{0.35\textwidth}%
\[
\delta=\frac{D_{T}}{D_{D}}
\]
\end{minipage} & %
\begin{minipage}[t]{0.4\textwidth}%
Diffusivity ratio.%
\end{minipage}\tabularnewline
\hline 
\begin{minipage}[t]{0.35\textwidth}%
\[
\tau=\frac{T_{T}}{\eta T_{D}}
\]
\end{minipage} & %
\begin{minipage}[t]{0.4\textwidth}%
Non--dimensional length of the egg deposition season of \emph{T. sinensis}.%
\end{minipage}\tabularnewline
\hline 
\begin{minipage}[t]{0.35\textwidth}%
\[
E_{T}=\gamma N_{T}
\]
\end{minipage} & %
\begin{minipage}[t]{0.4\textwidth}%
Number of female larvae per adult female of \emph{T. sinensis} that
survive the overwintering in optimal conditions.%
\end{minipage}\tabularnewline
\hline 
\end{tabular}
\par\end{centering}

\caption{\label{tab:Rescaled-parameters}Non--dimensional variables and parameters.
Here the unit of time is $T_{D}$ and the unit of space is $\sqrt{D_{D}T_{D}}$.}
\end{table}

\subsection{The value of the parameters\label{sub:The-value-of-the-parameters}}

The mathematical model developed in this section depends on 11 free
parameters, listed in Table \ref{tab:Parameters}. Of these, one depends
on the physiology and on the distribution of the chestnuts, namely
the bud density $\beta_{max}$. Its numerical value and its significance
will be discussed at the beginning of the next section.

The other 10 parameters are related to the physiology of either \emph{D.
kuriphilus} or to \emph{T. sinensis}. The value of 6 of these, namely
$M$, $a$, $T_{D}$, $T_{T}$, $N_{D}$, $N_{T}$, is fairly well-known;
the value of $\eta$ and $\gamma$ is debatable, and it might be different
in different regions of the world; the value of $D_{D}$ and $D_{T}$
is unknown, but the model links it to more readily measurable quantities.
We shall now briefly discuss all of them in turn.

The maximum number $M$ of eggs of \emph{D. kuriphilus} per chestnut
bud is only used in the definition of the non--dimensional densities
(see Table \ref{tab:Rescaled-parameters}) but it does not enter in
the parameters that appear in the non--dimensional model (\ref{eq:complete_model}).
Thus, any uncertainty in its value would not affect the dynamics.
Then there are three intervals of time: the life span $a$ of adult
individuals of \emph{D. kuriphilus}; the number of days $T_{D}$ during
which the adults of \emph{D. kuriphilus} are active (that is, the
length of what we have called the ``\emph{Dryocosmus} season'');
and the number of days $T_{T}$ during which the adults of \emph{T.
sinensis} are active (the ``\emph{Torymus} season''). What matters
for the model are the non--dimensional ratios $T_{D}/a$ and $T_{T}/T_{D}$.
We take 10 as the reference value for the first, and 1 for the second.
We have verified that any discrepancy from these reference values,
as long as it is compatible with the observational uncertainties,
makes little difference in the end results. In particular, in the
spatially homogeneous case, the calculations of the previous section
show that the map (\ref{eq:complete_map}) does not depend on their
ratio. Finally we have $N_{D}$ and $N_{T}$, respectively the average
number of eggs carried by \emph{D. kuriphilus} and \emph{T. sinensis}
females. For the first we take the reference value of 150 eggs per
female, and for the second we take 70 eggs per female. In the non--dimensional
model (\ref{eq:complete_model}) these parameters always appear multiplied,
respectively, by $\eta$ and $\gamma$. Any uncertainty in the value
of $N_{D}$ and $N_{T}$ is surely swamped by the uncertainty in these
two parameters.

In fact, the value of the two overwintering survival fractions $\eta$
(of \emph{D. kuriphilus}) and $\gamma$ (of \emph{T. sinensis}, which
also includes the sex ratio) are debatable. The works of \citet{CooperRieske07}
and of \citet{Piao92} suggest intermediate values for these parameters.
However, our own observations (published in \citealp{Quacchia13}
for \emph{D. kuriphilus} and yet unpublished for \emph{T. sinensis})
suggest much higher survival fractions. Whether these discrepancies
are due to regional variations (USA and Japan vs subalpine Europe)
or to some other cause is, at present, not known. Therefore, in the
following, we devote much attention to studying the dependence of
the dynamics on the value of the overwintering survival fractions.

The two diffusion coefficients $D_{D}$ and $D_{T}$, respectively
of \emph{D. kuriphilus} and \emph{T. sinensis}, are completely unknown.
In §\ref{sub:Space-dependent-dynamics} we estimate the value of $D_{D}$
on the basis of the model results and of the observed speed with which
a population of \emph{D. kuriphilus} is able to invade a chestnut
forest. Not enough data are available for attempting a similar deduction
with $D_{T}$. The effect of changing the diffusivity is studied in
detail in §\ref{sub:Space-dependent-dynamics-of-complete-model}.

\section{Results\label{sec:Results}}

\subsection{Space-independent dynamics\label{sub:Space-independent-dynamics}}

The map (\ref{eq:complete_map}), which describes the time evolution
of spatially homogeneous populations of \emph{D. kuriphilus} parasitized
by \emph{T. sinensis}, predicts that, starting from non-zero densities
of both species, the subsequent dynamics will continue to have non-zero
densities at all later years, with upper bounds determined by the
availability of buds (for \emph{D. kuriphilus}) and of galls (for
\emph{T. sinensis}; see sec. \ref{sub:Global-boundedness} for the
mathematical proof). This property alone, however, does not guarantee
the survival of either species. If, at some point in time, the modeled
egg density of a species drops to sufficiently low values, then the
model is predicting a local extinction of that species. An order--of--magnitude
estimate of the threshold density that signals extinction may be obtained
as follows: a full-grown chestnut tree in spring produces about $10^{4}$
buds; typical production orchards have a density of $100-200$ trees
per hectare, while coppices may have up to $1000$ stems per hectare,
but with less buds per stem than in individual trees \citep{Bounous14}.
Thus we have $\beta_{max}\approx2\cdot10^{6}$ $\mathrm{buds}\,\mathrm{ha}^{-1}$,
and, allowing for uncertainties in the above figures, it follows that
$V_{max}$ ranges between $10^{7}-10^{8}$ $\mathrm{eggs}\,\mathrm{ha}^{-1}$.
Therefore, non--dimensional densities $v_{n}$, $q_{n}$ below $10^{-7}-10^{-8}$
correspond to less than one insect per hectare. For an isolated, hectare-wide
orchard, this would be the extinction threshold. For a chestnut woodland
spanning several square kilometers the threshold would be proportionally
lower.

\begin{figure}
\begin{centering}
\includegraphics[width=0.85\textwidth]{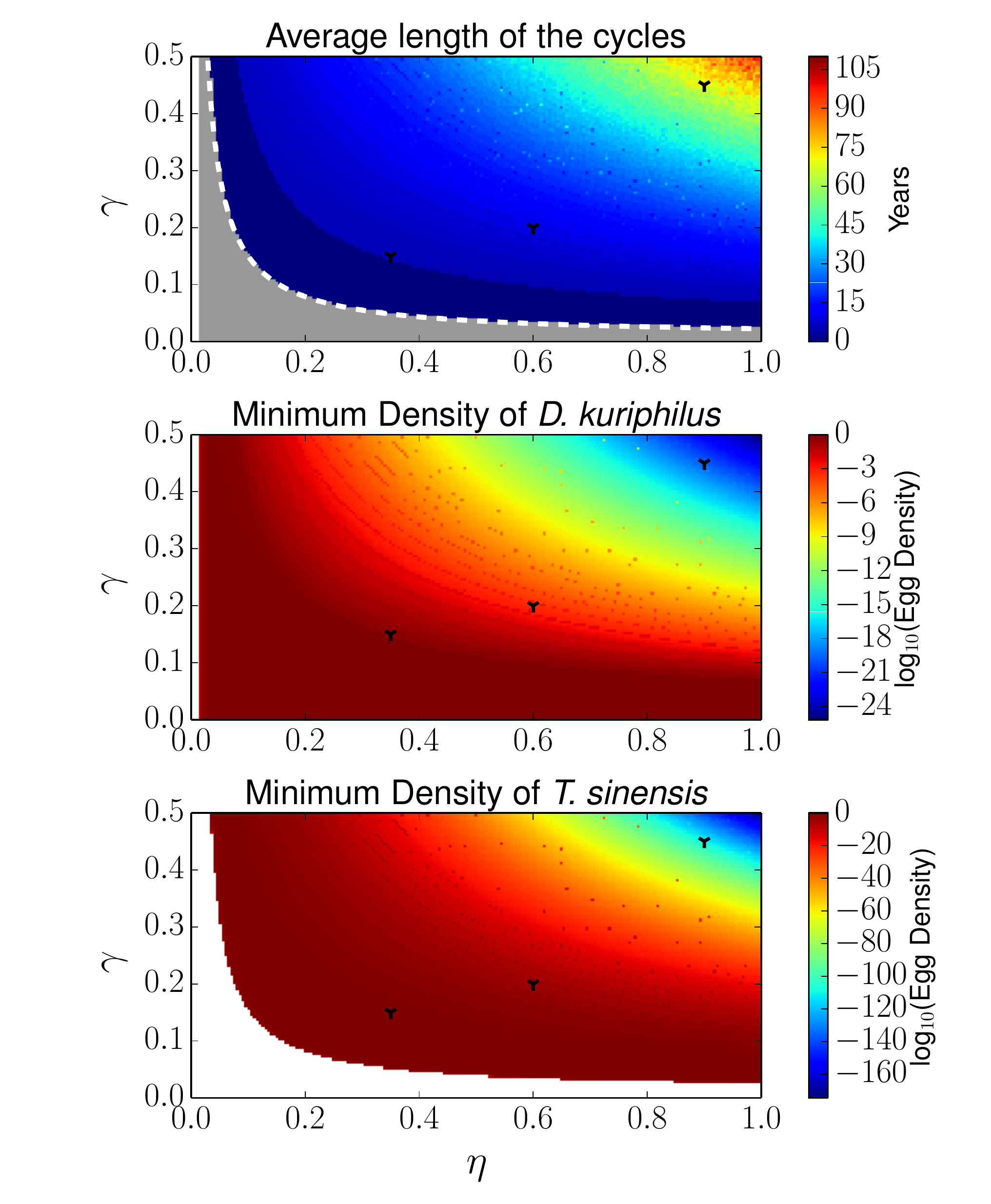}
\par\end{centering}

\caption{\label{fig:periods}As a function of the overwintering survival fractions
$\eta$ and $\gamma$, top panel: average length of the predator--prey
cycles; zero means that the coexistence fixed point is stable; the
gray area that \emph{T. sinensis} asymptotically becomes extinct;
the white area that both species asymptotically become extinct; the
white dashed line is the graph of (\ref{eq:gamma-treshold}). Middle
panel: $\log_{10}$ of the minimum density of \emph{D. kuriphilus}
eggs; the white region corresponds to asymptotic extinction. Lower
panel: $\log_{10}$ of the minimum density of \emph{T. sinensis} eggs;
the white region corresponds to asymptotic extinction. For each pair
$(\eta,\gamma)$ the statistics are computed over 3000 iterations
of the map, after a 2000 iterations transient. The black markers show
the parameters of Figures (\ref{fig:eta35_gamma15}), (\ref{fig:eta60_gamma20}),
(\ref{fig:eta90_gamma45}). Note that $\gamma$, the survival fraction
of \emph{T. sinensis,} also includes the sex ratio, and therefore
may not be greater than $0.5$.}
\end{figure}

\begin{figure}
\begin{centering}
\includegraphics[width=0.99\textwidth]{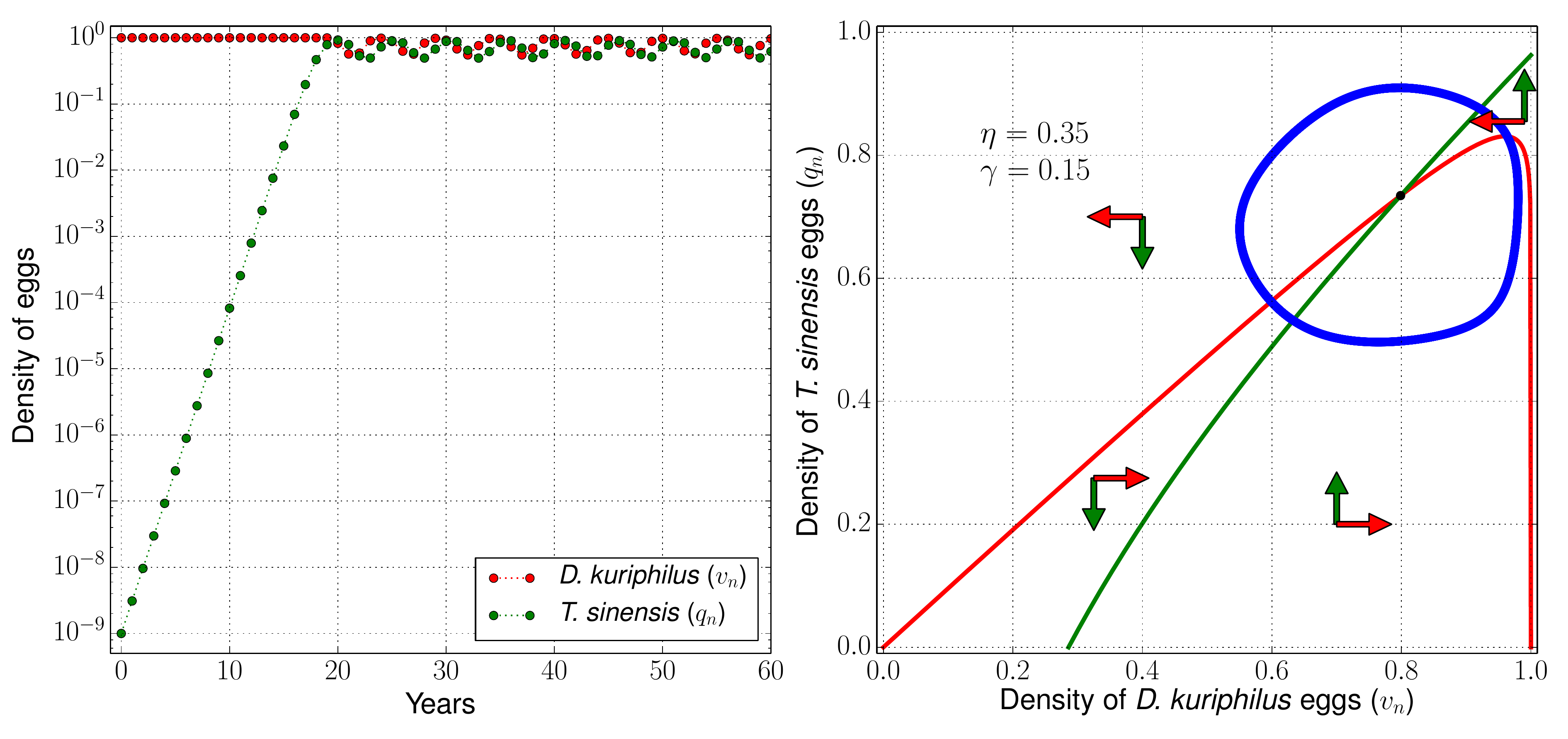}
\par\end{centering}

\caption{\label{fig:eta35_gamma15}Left panel: time evolution of \emph{D. kuriphilus}
(red dots) and \emph{T. sinensis} (green dots) egg density when the
fraction of (female) larvae surviving the overwintering is, respectively
$\eta=0.35$ and $\gamma=0.15$; see the text for the other parameters.
Right panel: the red and the green curves are, respectively, the $v-$nullcline
and the $q-$nullcline, which partition the plane in four regions;
their intersection, marked by the black dot, is the unstable coexistence
fixed point; the blue loops shows the states that the system occupies
for asymptotically large times; the red horizontal arrows and the
vertical green arrows show, for each of the four regions, whether
the densities of the next state will be larger of smaller than those
of a state in that region. }
\end{figure}

\begin{figure}
\begin{centering}
\includegraphics[width=0.99\textwidth]{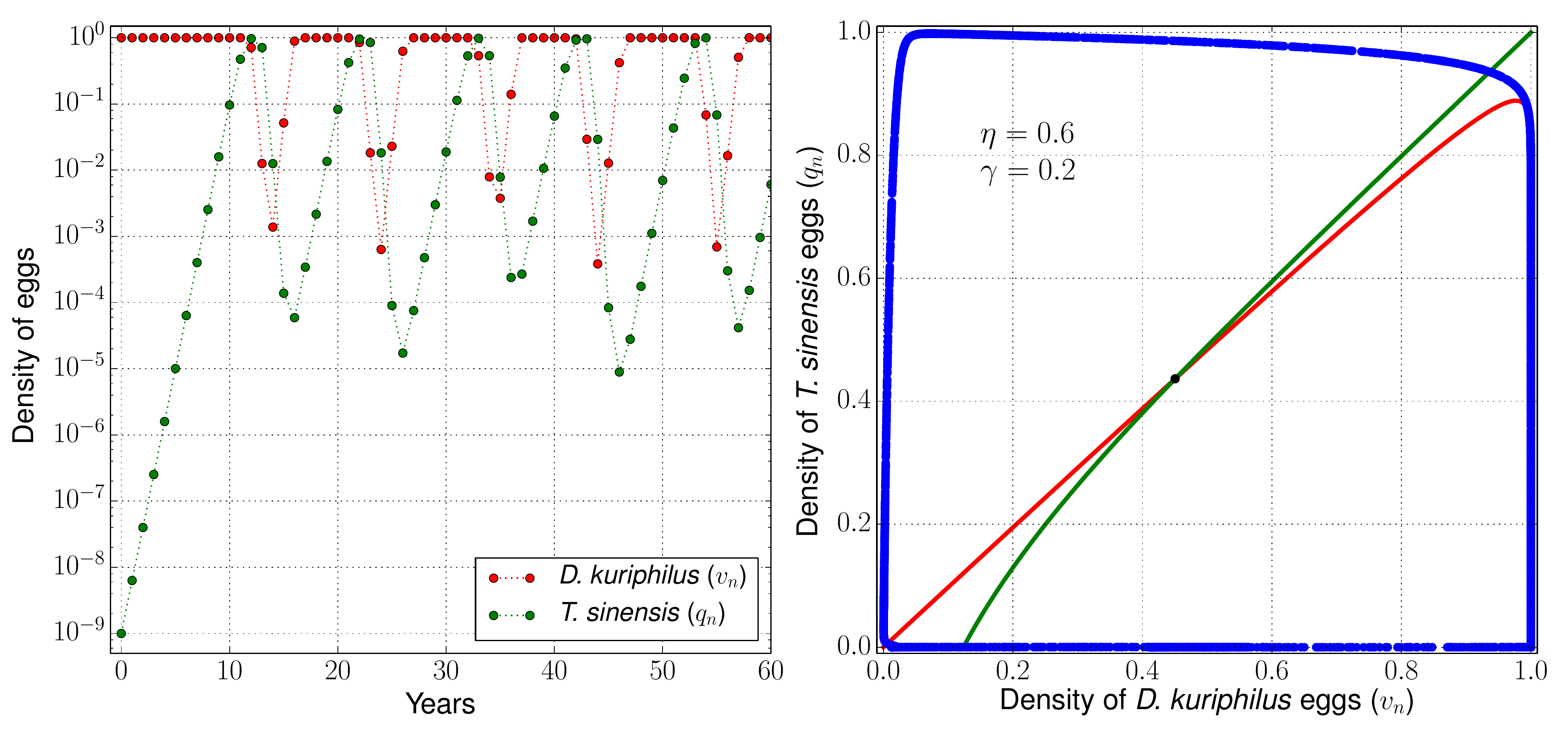}
\par\end{centering}

\caption{\label{fig:eta60_gamma20}As Figure \ref{fig:eta35_gamma15}, but
with $\eta=0.6$ and $\gamma=0.2$. These parameters might be representative
of a situation in which \emph{T. sinensis} suffers from hyperparasitism,
as it is hypothesized for Japan.}
\end{figure}

\begin{figure}
\begin{centering}
\includegraphics[width=0.99\textwidth]{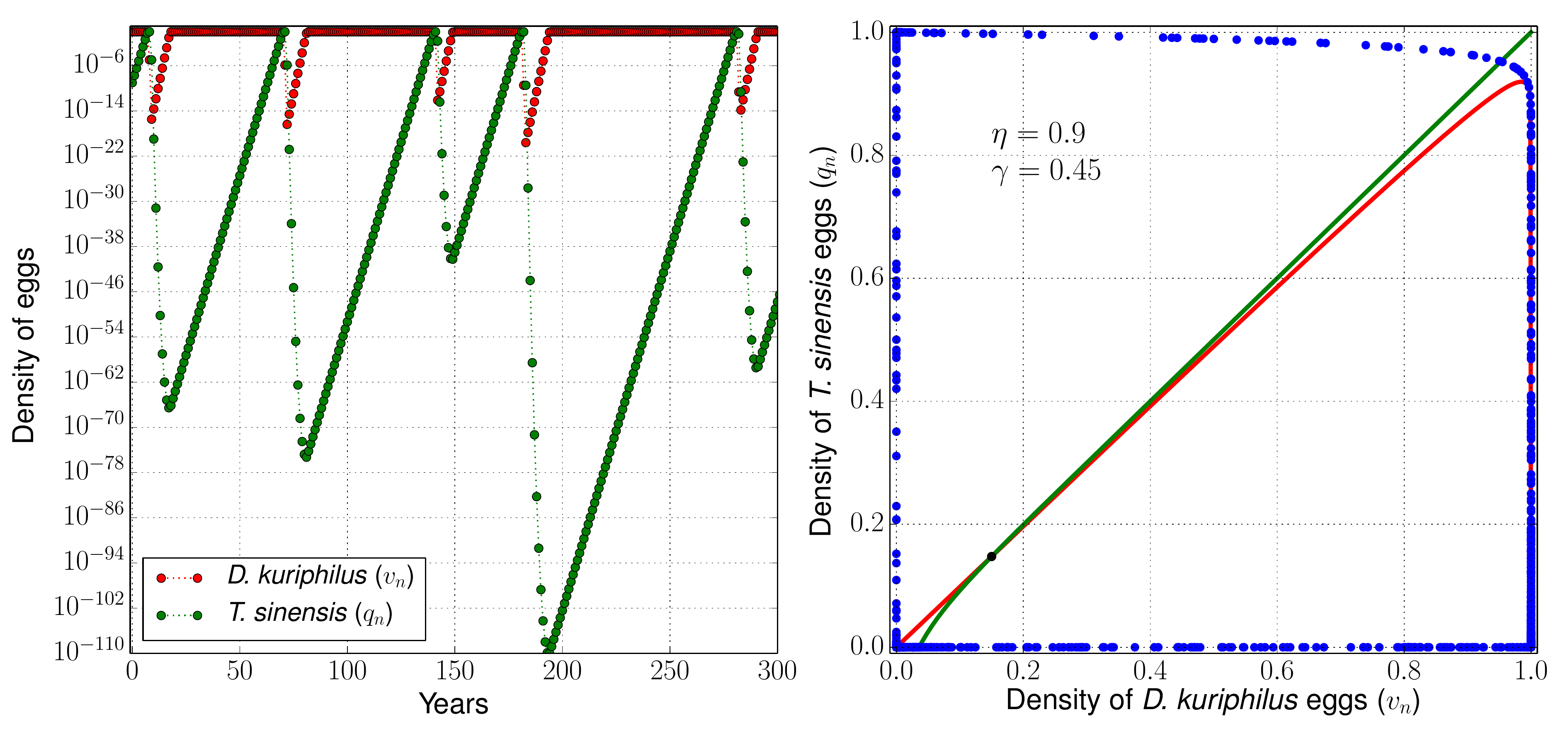}
\par\end{centering}

\caption{\label{fig:eta90_gamma45}As Figure \ref{fig:eta35_gamma15}, but
with $\eta=0.9$ and $\gamma=0.45$. These are parameters that we
consider to be realistic for Europe. Note the extreme excursion of
densities and the logarithmic density axis of the left panel.}
\end{figure}

The map (\ref{eq:complete_map}) depends on three parameters: $\eta$,
$E_{T}$, $k$. The last two, in turn, depend on other parameters,
namely $\gamma$, $N_{T}$, $\eta$, $N_{D}$ and $\mu$ (see Tables
\ref{tab:Parameters} and \ref{tab:Rescaled-parameters}). We shall
discuss the dynamics of the map as a function of the overwintering
survival fractions $\eta$ and $\gamma$ (owing to the uncertainty
of their value) and fix the other parameters to the following values:
$N_{D}=150$ (eggs per \emph{D. kuriphilus} adult), $N_{T}=70$ (eggs
per \emph{T. sinensis} female adult), $\mu=10$ (ratio of lengths
of the season and individual life span for \emph{D. kuriphilus}).
Uncertainties in the value of $\mu$ do not produce large changes:
going from $\mu=3$ to $\mu=20$ gives about 15\% difference in the
value of the constant $k_{-}$ in (\ref{eq:Skellam_constant}). In
Figures \ref{fig:periods} to \ref{fig:eta90_gamma45} we use $k=k_{-}$,
because this choice gives a better approximation at low densities.
Note that using $k=k_{+}$, is equivalent to using a larger value
of $E_{D}$ with $k=k_{-}$. In order to avoid inaccuracies due to
numerical cancellation errors, all calculations for producing the
figures have been carried out with 200 decimal significant digits,
using an arbitrary precision numerical library \citet{Mpmath}. A
more general and technical analysis of the map is given in Appendix
\ref{sub:Mathematical-properties-of-map}. 

Depending on the values of $\eta$ and $\gamma$ there are 4 possible
dynamical outcomes qualitatively distinct: extinction of both species,
extinction of the parasitoid, steady coexistence, and predator--prey
cycles. The first 3 occur for unrealistically low values of these
parameters. If $\eta$ is as low as to make $k<1$ in (\ref{eq:complete_map}),
then the egg density of \emph{D. kuriphilus} asymptotically goes to
zero. As the gall wasp goes extinct, so does \emph{T. sinensis}, for
lack of galls where to lay eggs. This region of the parameter space
is represented by the white vertical strip in Figure \ref{fig:periods}
(top panel). If $\eta$ is such that $k>1$ and $\gamma$ is sufficiently
low, then only \emph{T. sinensis} becomes extinct, and \emph{D. kuriphilus}
reaches the non-zero fixed point of Skellam's map (\ref{eq:Skellam-model}).
This is the gray region in Figure \ref{fig:periods} (top panel).
The exact threshold value of $\gamma_{tr}$ cannot be expressed in
simple terms, but a good approximation (the white dashed line in Figure
\ref{fig:periods}, top panel) is 
\begin{equation}
\gamma_{tr}\approx\frac{1}{N_{T}\left(1-e^{-\eta}\right)}.\label{eq:gamma-treshold}
\end{equation}

The dark blue region above the threshold in Figure \ref{fig:periods}
(top panel) corresponds to the survival fractions at which both species
survive and reach a stable fixed point. The shape of this region shows
that, according to the model, a steady coexistence of both species
may only occur if the overwintering survival fraction of at least
one of the two species is unrealistically low. 

When it exists, we find a unique coexistence fixed point. It can be
visualized as the intersection between the set of points $(v_{n},q_{n})$
such that $q_{n+1}=q_{n}$ (the green lines in the right panel of
Figures \ref{fig:eta35_gamma15}, \ref{fig:eta60_gamma20}, \ref{fig:eta90_gamma45})
and the set of points $(v_{n},q_{n+1})$ such that $v_{n+1}=v_{n}$
(the red lines in the right panel of Figures \ref{fig:eta35_gamma15},
\ref{fig:eta60_gamma20}, \ref{fig:eta90_gamma45}). We shall call
these sets, respectively, $q-$nullcline and $v-$nullcline, and they
intersect, at most, at a single point (see Appendix \ref{sub:Nullclines}
for details). If the overwintering survival fractions $\eta$ and
$\gamma$ in Figure \ref{fig:periods} (top panel) lie beyond the
dark blue region of steady coexistence, a coexistence fixed point
still exists, but is unstable, therefore the insect egg densities
fluctuate from year to year. When $q_{n}$ is above the $q-$nullcline,
then $q_{n+1}<q_{n}$, and if $q_{n}$ is below, then $q_{n+1}>q_{n}$
. Analogously, when $v_{n}$ is above (below) the $v-$nullcline,
then $v_{n+1}$ is smaller (larger) than $v_{n}$. These drop or raise
tendencies are depicted in the left panel of Figure \ref{fig:eta35_gamma15},
by the green vertical arrows for $q_{n}$, and by the red horizontal
arrows for $v_{n}$. The arrows in the right panel of Figure \ref{fig:eta35_gamma15}
suggest that the sequence of states in a $q_{n}$ vs $v_{n}$ diagram
rotates anticlockwise around the unstable fixed point, corresponding
to cyclical increases and decreases of both species, in which maxima
and minima of \emph{T. sinensis} follow the maxima and minima of \emph{D.
kuriphilus}, yielding the kind of fluctuations that are ubiquitous
in predator-prey dynamics \citep[e.g. ][ch. 5]{TheoreticalEcology2007}.
For generic values of $\eta$ and $\gamma$ these fluctuations are
not periodic, nor asymptotically approach a periodic oscillation.
However the cycles are characterized by a fairly well-defined time
scale. 

Figure \ref{fig:eta35_gamma15} (left panel) shows 60 years of dynamics
that one obtains by adopting for the overwintering survival fractions
the very low values $\eta=0.35$ and $\gamma=0.15$, located just
beyond the steady coexistence region. The initial state is designed
to simulate the release of a tiny amount of \emph{T. sinensis} in
a large chestnut forest infested by \emph{D. kuriphilus}. Thus we
take $v_{n}=1$ and $q_{n}=10^{-9}$. During the first two decades
the population of \emph{T. sinensis} grows steadily from the very
low initial density, while the population of \emph{D. kuriphilus}
remains essentially unaffected by the presence of the parasitoid.
When $q_{n}$ approaches 1, then $v_{n}$ begins to decline. From
then on, the densities of the two species oscillate in cycles that
are about 5 years long. Omitting the initial transient, and plotting
$q_{n}$ vs $v_{n}$ one finds that the succession of states describes
the blue loop depicted in Figure \ref{fig:eta35_gamma15} (right panel).
With these low survival fractions the loop winds relatively close
to the unstable fixed point (the black dot at the intersection of
the nullclines). Note that with these parameters the biological control
is not achieved in a satisfactory way: the egg density of \emph{D.
kuriphilus} never drops below 50\% of its initial value.

In Figure \ref{fig:eta60_gamma20} we show the dynamics when the overwintering
survival fractions are increased to $\eta=0.6$ and $\gamma=0.2$.
Albeit still low, these values might represent the case of Japan,
where, according to \citet{Murakami91}, a large amount of non--specialist
parasitoid and hyperparasitoids species, assumed to be always present
in the environment at a roughly constant concentration, cause a high
overwintering mortality both in \emph{D. kuriphilus}, and, even more,
in \emph{T. sinensis}. The left panel of Figure \ref{fig:eta60_gamma20}
shows that the transient growth of \emph{T. sinensis} is shortened
to about a decade, after which it starts to dent the population of
\emph{D. kuriphilus}. The cycles after the transient have a length
of 10--11 years, which is a time scale that roughly matches the observations
in Japan (Moriya, personal communication). In the model, the egg concentration
of \emph{D. kuriphilus} remains almost constant, and very close to
1, for 6--7 years, while the density of \emph{T. sinensis} grows.
Then, in the turn of 1--2 years \emph{T. sinensis} peaks and causes
a sudden drop in the concentration of \emph{D. kuriphilus}, and consequently,
also the population of \emph{T. sinensis} drops in the following years.
The recovery of \emph{D. kuriphilus} occurs in 1--2 years, starting
from minimum densities that may be smaller than $10^{-3}$. The population
of \emph{T. sinensis} continues to drop until the recovery of \emph{D.
kuriphilus} is almost complete, then it starts to increase. By this
time the density of \emph{T. sinensis} may have reached densities
almost as low as $10^{-5}$. The decline and the subsequent recovery
of \emph{T. sinensis} span almost all the length of the cycle. High
densities of \emph{T. sinensis} occur only for 2--3 years in each
cycle. The right panel of Figure \ref{fig:eta60_gamma20} shows the
succession of states (in blue, the initial transient was omitted)
looping anticlockwise around the unstable fixed point (the black dot).
In this case the loop is pushed much further away from the fixed point
than in Figure \ref{fig:eta35_gamma15}, and the densities $v_{n}$,
$q_{n}$ almost always assume either very low values or values close
to 1. Even in this case a satisfactory biological control is not achieved. 

This kind of dynamics, in which \emph{D. kuriphilus} remains most
of the times at densities close to 1, interrupted by brief bursts
in the population of \emph{T. sinensis}, rather than performing mild
oscillations at intermediate values, occurs every time that the overwintering
survival fractions are significantly removed from the stability region
of the fixed point. For example, if the overwintering survival fractions
are increased to values that we consider realistic for Europe, such
as $\eta=0.9$ and $\gamma=0.45$, we observe the same stasis and
burst dynamics as in the previous case, but with much longer cycles,
that may last several decades (Figure \ref{fig:eta90_gamma45}). However,
the really remarkable feature of this case is the extreme depth of
the drops in the population density of both species. The left panel
in Figure \ref{fig:eta90_gamma45} shows that when\emph{ }the egg
density of \emph{T. sinensis} becomes close to 1, then the egg density
of \emph{D. kuriphilus}, in a single season, drops to values that
may be smaller than $10^{-20}$. The subsequent recovery of \emph{D.
kuriphilus} is not short, but requires several years, during which
the population of \emph{T. sinensis,} for lack of deposition sites,
decreases to absurdly low values. As we argued at the beginning of
this section, cycles that reach minima this low are a mathematical
fiction. In reality, the model is stating that \emph{T. sinensis},
after the initial transient, wipes out the local population of \emph{D.
kuriphilus}, and then becomes extinct itself. We would like to stress
that this is really a fifth dynamical regime, and it should not be
confused with the extinction that takes places at the opposite end
of the parameter space, when the overwintering survival fractions
are very close to zero. 

The three panels of Figure \ref{fig:periods} show, from top to bottom,
the average length of the cycles, and the minimum density attained
during a cycle for \emph{D. kuriphilus} (middle panel) and \emph{T.
sinensis} (lower panel) as a function of $\eta$ and $\gamma$. Note
that all reasonably high values of the survival fractions yield long
cycle lengths and extremely low densities at minimum, signaling the
extinction of both species. However, this does not necessarily means
that the map is forecasting a successful biological control. Because
the minima of \emph{T. sinensis} are generally much lower than those
of \emph{D. kuriphilus}, there is the possibility that \emph{T. sinensis}
reaches extinction--level densities before \emph{D. kuriphilus}, which
would then remain completely unchecked.

Finally, we mention that, for selected values of $\eta$ and $\gamma$,
the map (\ref{eq:complete_map}) appears to be characterized by periodic
cycles with amplitude and length smaller than those found at different,
but very close, values of the parameters (these are the scattered
dots of color slightly different than the surroundings visible in
Figure \ref{fig:periods}). These \emph{regularity windows} are commonplace
in discrete-time dynamical systems such as the map (\ref{eq:complete_map})
and are unlikely to persist if subject to the random perturbations
that are always present in a natural environment, but are absent in
this simple model. Thus their presence does not change the overall
qualitative description of the dynamics given above.

\subsection{Space-dependent dynamics of \emph{D. kuriphilus}\label{sub:Space-dependent-dynamics}}

Before discussing the complete model (\ref{eq:complete_model}) it
is appropriate to analyze the dynamics of \emph{D. kuriphilus} alone,
as it invades an idealized forest. We shall consider a 1--dimensional
spatial domain, that could be thought of as a very long strip of trees
whose width is negligible with respect to its length. 

\begin{figure}
\begin{centering}
\includegraphics[width=0.99\textwidth]{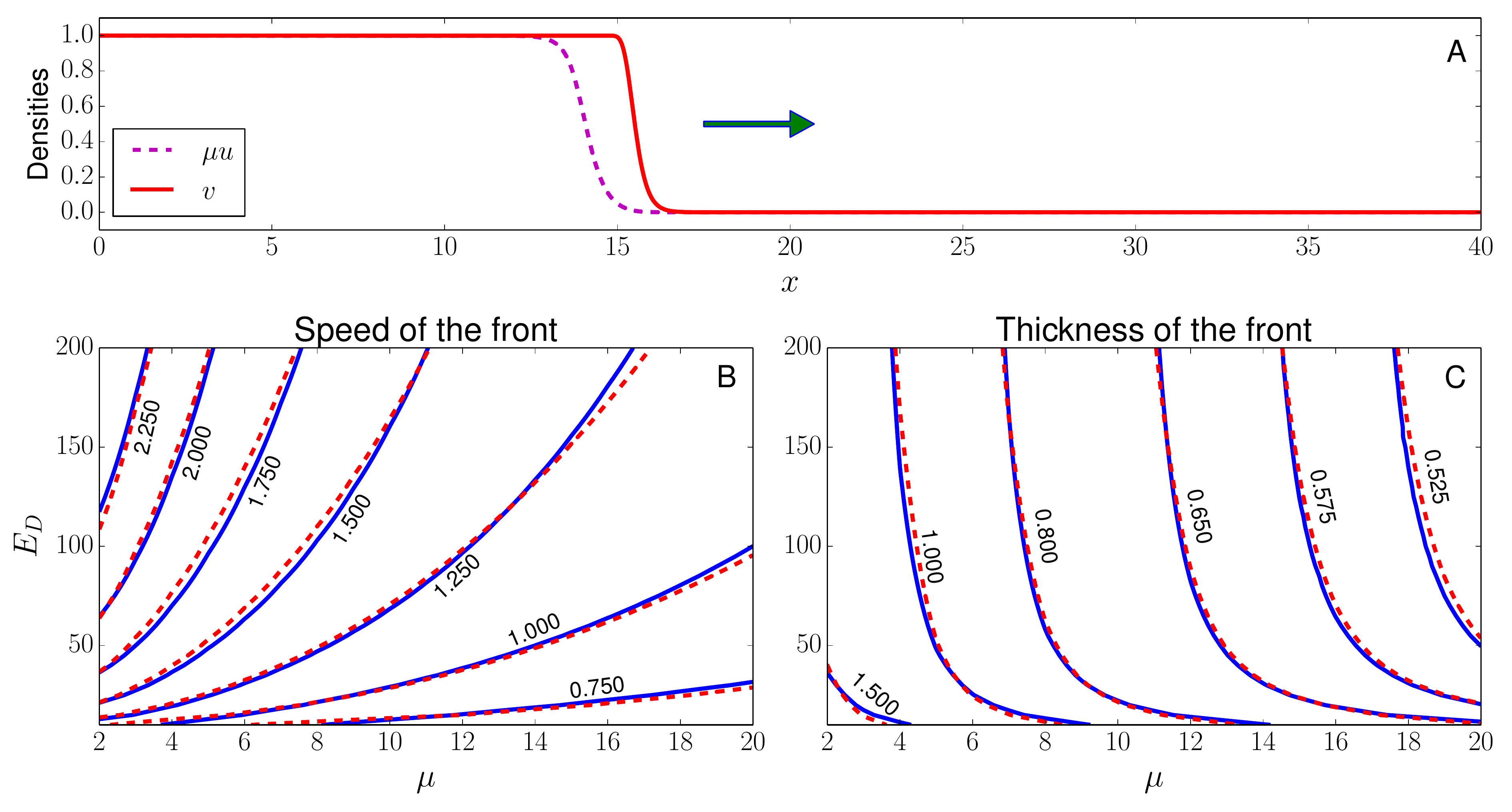}
\par\end{centering}

\caption{\label{fig:cinipide_speed_thickness}A) Density of eggs (continuous
line) and of adults (dashed line) of \emph{D. kuriphilus} at the end
of the 10th season in a numerical solution of equations (\ref{eq:cinipide_model_uv_rescaled})
with $\mu=10$ and $E_{D}=135$, where the pest is initially introduced
at the left end of the one-dimensional domain. For clarity the density
of adults is multiplied by $\mu$. The arrow shows the direction of
propagation of the front. B) Contour lines of the speed of the front
as a function of the parameters $\mu$ and $E_{D}$. The solid lines
are the numerical results, and the dashed lines are the fit $(1.95(E_{D}-8)^{1/5})/(\mu+2.75)^{1/2}$.
C) Contour lines of the thickness of the egg density front as a function
of the parameters $\mu$ and $E_{D}$. The solid lines are the numerical
results, and the dashed lines are the fit $((2.26E_{D}+7)/E_{D})/(\mu+0.1E_{D}^{1/2})^{1/2}$.
All quantities in these figure use the non--dimensional units of Table
\ref{tab:Rescaled-parameters}.}
\end{figure}

If \emph{D. kuriphilus} is released at one end of the strip, in the
absence of \emph{T. sinensis}, the equations (\ref{eq:the_cinipide_model}),
and their non--dimensional counterpart (\ref{eq:cinipide_model_uv_rescaled}),
describe the propagation of the population of the pest as it invades
the domain. This is a traveling front joining the region in which
the forest is fully infested by the pest to the region in which the
pest is still absent, as illustrated in Figure \ref{fig:cinipide_speed_thickness}A.
Note that, owing to the large number of eggs that can be laid by a
single individual, a relatively small density of adults is sufficient
to saturate all of the available buds. Therefore, at the end of the
season, the egg density front is offset with respect to the density
front of the adult population. Numerical simulations (we used centered,
second--order, finite--differences discretization for the diffusion
term and Heun's method for timestepping) show a strong analogy with
the propagation of a burning front, and the solutions are reminiscent
of those of the well--known Kolmogorov--Fisher equation, which is
the prototypical example for this kind of phenomena (\citealp[see e.g.][ § 13.2]{Murray07}).
For the K-F equation a simple argument based on dimensional analysis
shows that the speed and thickness of the front are directly proportional
to the square root of the diffusion coefficient. The thickness is
also directly proportional to the characteristic time of the chemical
reactions, but the speed is inversely proportional to it.

Equations (\ref{eq:the_cinipide_model}) are more complicated. The
change of variables that brings (\ref{eq:the_cinipide_model}) into
(\ref{eq:cinipide_model_uv_rescaled}) also suggests a proportionality
of speed and thinkness of the gall wasp front to the square root of
the diffusivity. However, there are three distinct time scales that
characterize the reaction--like terms of equations (\ref{eq:the_cinipide_model}):
the season length $T_{D}$, the individual life span $a$, and the
reciprocal of the egg deposition rate, $a/N_{D}$. Thus, in the non--dimensional
equations (\ref{eq:cinipide_model_uv_rescaled}) there remain two
independent parameters, namely $E_{D}$ and $\mu$. Figures \ref{fig:cinipide_speed_thickness}B
and \ref{fig:cinipide_speed_thickness}C show the dependence of speed
and thickness of the front on $E_{D}$ and $\mu$ in a wide range
of values. These data (represented by the solid lines) use the non--dimensional
units of Table \ref{tab:Rescaled-parameters}. In particular, the
unit of length is $\sqrt{D_{D}T_{D}}$ and the unit of speed is $\sqrt{D_{D}/T_{D}}$.
The speed of the front is evaluated as the speed of the point where
the non--dimensional egg density $v$ is equal to 1/2. The thickness
of the front is estimated as $(\partial v/\partial x)^{-1}$, where
the derivative is evaluated at the same point. Both quantities are
computed from the results of numerical solutions of the equations
(\ref{eq:cinipide_model_uv_rescaled}) subject to the conditions (\ref{eq:Cinipede_initial_conditions})
and (\ref{eq:BC_no_U_flux}). The data may be fitted reasonably well
with simple analytic expressions (dashed lines, see the figure captions
for their expression in terms of $E_{D}$ and $\mu$). In terms of
dimensional variables, and of the parameters of Table \ref{tab:Parameters},
the fits for the speed $S$ and the thickness $\Delta$ of the front
read 
\begin{equation}
S=1.95\left(\eta N_{D}-8\right)^{1/5}\sqrt{\frac{D_{D}}{T_{D}\left(\frac{T_{D}}{a}+2.75\right)}},\label{eq:cinipide_speed}
\end{equation}
\begin{equation}
\Delta=\left(\frac{2.26\eta N_{D}+7}{\eta N_{D}}\right)\sqrt{\frac{T_{D}D_{D}}{\frac{T_{D}}{a}+0.1\sqrt{\eta N_{D}}}}.\label{eq:cinipide_thickness}
\end{equation}
These expressions are not formally deduced from the equations (we
postpone this issue to a future work), and thus should be considered
to be reliable only within the parameter range of Figures \ref{fig:cinipide_speed_thickness}B
and \ref{fig:cinipide_speed_thickness}C. Nevertheless, they are fully
satisfactory for the problem of determining the magnitude of the diffusion
coefficient $D_{D}$ on the basis of the observed propagation speed
of the pest. Taking into account that each year the gall wasp is active
and mobile only during the interval of time $T_{D}$, the speed of
the front can also be expressed as $S=L/T_{D}$, where $L$ is the
length traveled in one year by the infestation (as reported by \citealp{EFSA10}). 

Taking $T_{D}=40$ d, $a=4$ d, $\eta=0.9$, $N_{D}=150$, from (\ref{eq:cinipide_speed})
we find
\begin{equation}
D_{D}\approx\frac{L^{2}}{83}\label{eq:D_D}
\end{equation}
where the denominator is expressed in days. Using this in (\ref{eq:cinipide_thickness})
we can link the thickness of the front to the length it travels in
a season, finding
\begin{equation}
\Delta\approx\frac{L}{2.1}.\label{eq:Delta_vs_L}
\end{equation}
For example, with $L=8\,\mathrm{km}$, we have a thickness of the
front $\Delta\approx3.8\,\mathrm{km}$, and a diffusion coefficient
$D_{D}\approx0.77\,\mathrm{km^{2}}\mathrm{d}^{-1}$. With a numerical
value for the diffusion coefficient we can explicitly convert our
non--dimensional lengths in kilometers, finding that, in this example,
one unit of length is $\sqrt{D_{D}T_{D}}\approx5.5\,\mathrm{km}$.

Assuming that the trajectories of individual insects approximate a
Brownian motion, Einstein's formula \citep[see, e.g., ][§1.2]{Gardiner04}
suggests that the typical displacement $l$ of an adult after $t$
days would be $l=\sqrt{2\, n\, D_{D}\, t}$, where $n$ is the dimensionality
of the domain. Thus, in our idealized 1--dimensional strip of trees
the displacement over the entire adult life span (4 days) would be
$l\approx2.5\,\mathrm{km}$, and in a 2--dimensional domain, such
as a real wood, it would be $l\approx3.5\,\mathrm{km}$.

\subsection{Space-dependent dynamics of the host--parasitoid system\label{sub:Space-dependent-dynamics-of-complete-model}}

\begin{figure}
\begin{centering}
\includegraphics[width=0.99\textwidth]{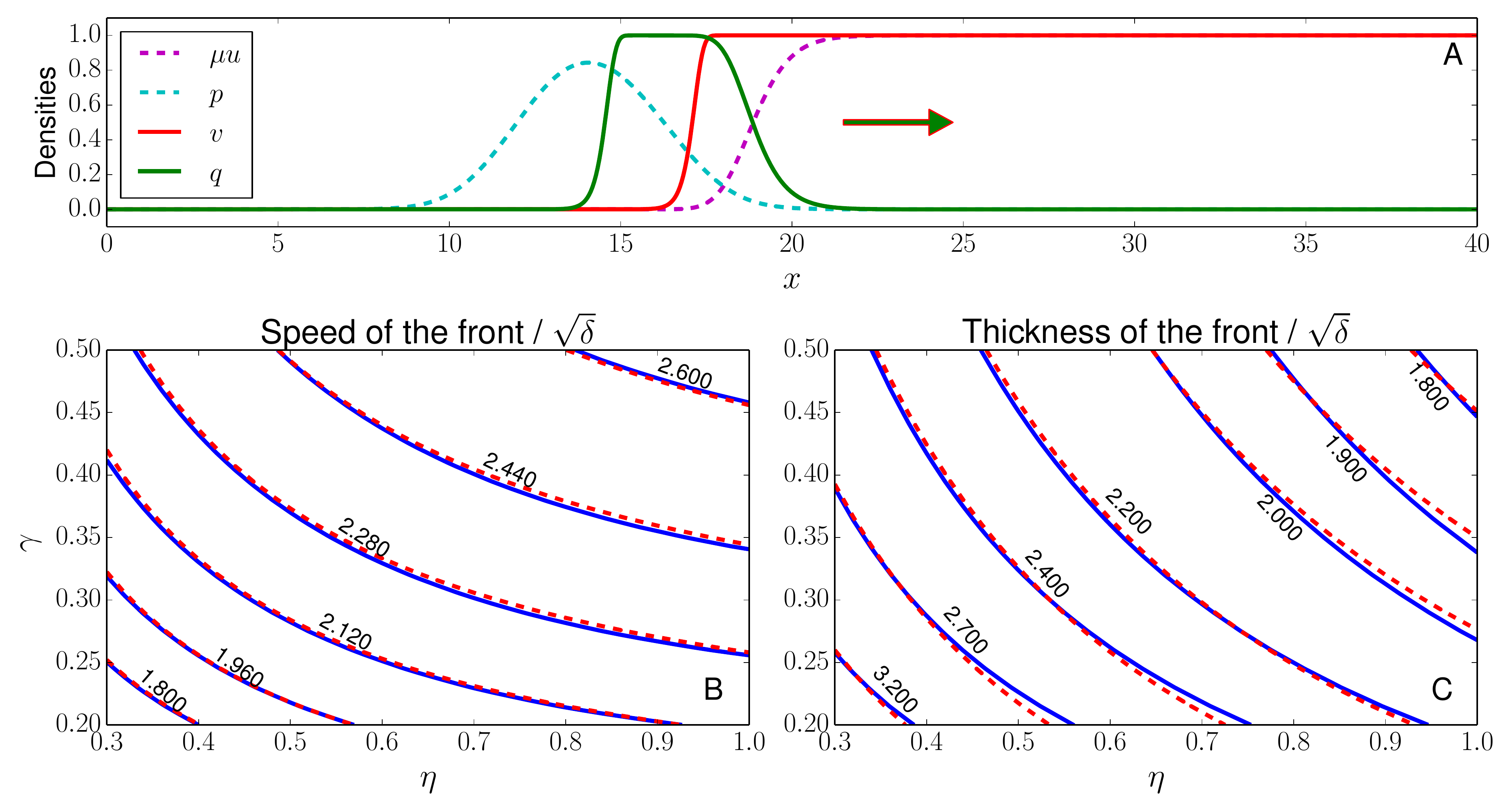}
\par\end{centering}

\caption{\label{fig:torymus_cinipide_speed_thickness}A) Density of eggs (red
continuous line) and of adults (magenta dashed line) of \emph{D. kuriphilus}
density of eggs (green continuous line) and adults (cyan dashed line)
at the end of the 7th season in a numerical solution of equations
(\ref{eq:complete_model}) with $\mu=10$, $E_{D}=135$, $E_{T}=31.5$,
$\gamma=0.45$, $\eta=0.9$, $\delta=1$, $\tau=1/\eta$, where the
pest is initially homogeneously distributed throughout the one--dimensional
domain, and the parasitoid is present only at its left end. For clarity
the density of adults of \emph{D. kuriphilus} is multiplied by $\mu$.
The arrow shows the direction of propagation of the front. B) Contour
lines of the speed of the front as a function of the overwintering
survival fractions $\eta$ and $\gamma$. The solid lines are the
numerical results, and the dashed lines are the fit $(\eta/(0.301\eta+0.021))(\gamma-0.05/(\eta+0.12))^{1/5}$.
C) Contour lines of the thickness of the egg density front as a function
of the overwintering survival fractions $\eta$ and $\gamma$. The
solid lines are the numerical results, and the dashed lines are the
fit $((0.88\eta+1.55)/(0.6+\eta))(\gamma-(1.-0.6\eta)/(17\eta))^{-1/5}$.
For values of $\delta$ different from 1, an excellent fit is obtained
by multiplying these expressions by $\sqrt{\delta}$. All quantities
in these figure use the non--dimensional units of Table \ref{tab:Rescaled-parameters}.}
\end{figure}

If we start from an initial condition in which the idealized 1--dimensional
forest is fully infested by the pest, and the parasitoid is introduced
at its left end, then, in the course of years, the parasitoid population
will propagate rightward, as depicted in Figure \ref{fig:torymus_cinipide_speed_thickness}A.
As the parasitoid propagates rightward, it causes a severe drop in
the population density of the pest, which develops a left--facing
region of high gradient, connecting the part of the forest which has
not yet been reached by the parasitoid, and thus is still fully infested,
to the part already swept by the parasitoid, where the pest density
has been severely reduced. The reduction in the pest density is mirrored
by a corresponding reduction of the parasitoid density, which faces
a drastic scarcity of its host in the region of the forest that has
already been swept. Therefore, the parasitoid population propagates
into an infested forest as a moving peak, rather than as a moving
kink.

The results of the numerical simulations show that the speed of propagation
of the parasitoid, and the thickness of the right--facing gradient
region of its egg density, are proportional to the square root of
the diffusivity ratio $\sqrt{\delta}$ (see Table \ref{tab:Rescaled-parameters}),
with excellent approximation, at least in the interval $\delta\in[0.1,10]$.
We have also computed the dependence of speed and thickness on the
overwintering survival fractions $\eta$ and $\gamma$. The results
are reported in Figures \ref{fig:torymus_cinipide_speed_thickness}B
and \ref{fig:torymus_cinipide_speed_thickness}C. These results may
be fitted by simple expressions, which, in terms of the parameters
of Table \ref{tab:Parameters}, read: 
\begin{equation}
S_{T}=\frac{\eta}{0.301\eta+0.021}\left(\gamma-\frac{0.05}{\eta+0.12}\right)^{1/5}\sqrt{\frac{D_{T}}{D_{D}}},\label{eq:Torymus_front_speed}
\end{equation}
\begin{equation}
\Delta_{T}=\frac{0.88\eta+1.55}{0.6+\eta}\left(\gamma-\frac{1.-0.6\eta}{17\eta}\right)^{-1/5}\sqrt{\frac{D_{T}}{D_{D}}}\label{eq:Torymus_front_thickness}
\end{equation}
These expressions are valid when the other parameters are $\mu=10$,
$E_{D}=135$, $E_{T}=31.5$, $\tau=1/\eta$, which should represent
fairly well the relevant physiological parameters of both the pest
and of the parasitoid, as discussed in §\ref{sub:The-value-of-the-parameters}. 

The density to which both the pest and the parasitoid drop on the
left--hand side of the right--moving peak, depends on the value of
the overwintering survival fractions $\eta$ and $\gamma$, roughly
in the same way as shown in Figure \ref{fig:periods} for the spatially
homogeneous case discussed in §\ref{sub:Space-independent-dynamics}.
At low and intermediate survival rates (such as those of Figures \ref{fig:eta35_gamma15}
and \ref{fig:eta60_gamma20}) the density drop spans at most a few
orders of magnitude, and it is thus insufficient to justify hopes
of eradication of the pest. 

At higher survival rates the severity of the density drop is as large
as to amply justify claims of local extinction: as the parasitoid
sweeps the forest, virtually no host will be left unparasitized. Unfortunately,
this effect alone does not guarantee a successful biological control.
In fact, our model shows cases in which the pest is able to recolonize
the empty forest left back by the passage of the parasitoid.

\begin{figure}
\begin{centering}
\includegraphics[width=0.5\textwidth]{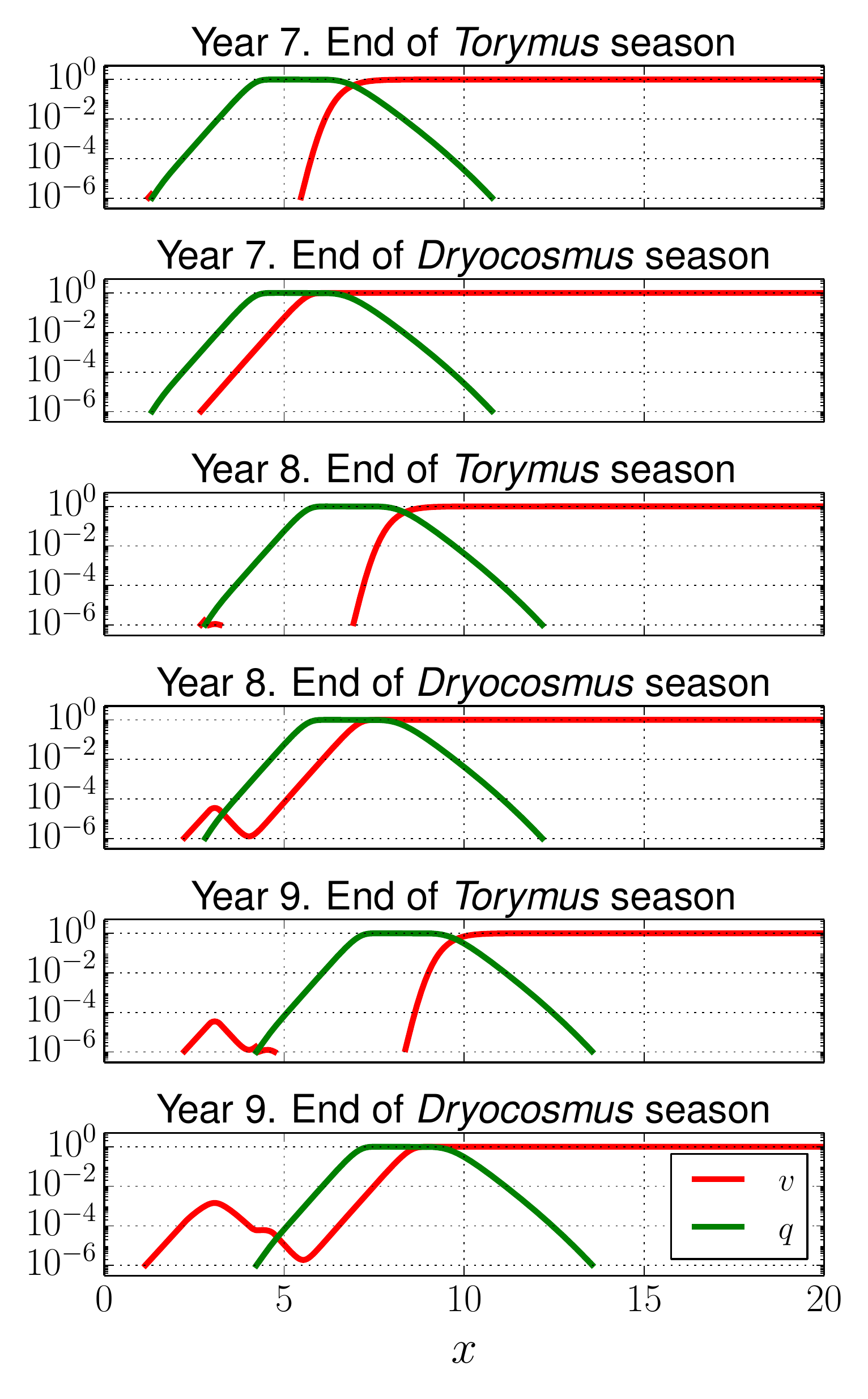}
\par\end{centering}

\caption{\label{fig:recolonization}A time sequence showing the beginning of
the recolonization of the forest by the pest after the passage of
the parasitoid. The red line is the density of eggs or unparasitized
larvae of the pest \emph{D. kuriphilus}; the green line is the density
of eggs or larvae of the parasitoid \emph{T. sinensis}. Note the logarithmic
scale. The ratio of the diffusivities (see Table \ref{tab:Rescaled-parameters})
is $\delta=0.3$, the other parameters are the same as in Figure \ref{fig:torymus_cinipide_speed_thickness}A. }
\end{figure}

Figure \ref{fig:recolonization} illustrates this phenomenon. In order
to demonstrate that the model really allows for cases of recolonization,
rather than failures of the parasitoid to attain a complete local
eradication of the pest, in the run that produced Figure \ref{fig:recolonization},
at the end of the \emph{Torymus} and of the \emph{Dryocosmus} seasons,
the egg density of both species was set to zero anywhere it was found
to be below a threshold of local extinction equal to $10^{-6}$ non--dimensional
units. The rationale of identifying areas of very low density in the
model as areas where no individual insects are likely to be present
was discussed at the beginning of §\ref{sub:Space-independent-dynamics}.
In this run the initial conditions correspond to a release of a small
amount of parasitoid in a region spanning 1 non--dimensional units
on the left end of the idealized 1--dimensional forest saturated by
the pest. In a few years the population of the parasitoid grows and
spreads rightward into the forest, locally wiping out the pest, and
leaving a region devoid of both host and parasitoid behind its passage.

The time sequence of Figure \ref{fig:recolonization} begins 7 years
after the release of the parasitoid. At the end of the \emph{Torymus}
season, little or no \emph{Dryocosmus} larvae remain unparasitized
in correspondence of the peak density of the parasitoid (Figure \ref{fig:recolonization},
panel ``Year 7 - End of \emph{Torymus} season''). Then the surviving
larvae of \emph{Dryocosmus} emerge, and diffuse in the forest, looking
for deposition sites. By the end of the season, much of the ground
lost to \emph{Torymus} is recovered by \emph{Dryocosmus}, that arrives
to lay some eggs even in the region on the left of the \emph{Torymus}
peak, where the presence of the parasitoid is dwindling because of
the scarcity of the host. Thus, at the end of year 7, on the left
of the \emph{Torymus} peak, both host and parasitoid are present,
and, moving leftward, their density declines at a similar rate (Figure
\ref{fig:recolonization}, panel ``Year 7 - End of \emph{Dryocosmus}
season''). The next year \emph{Torymus} once again wipes out all
\emph{Dryocosmus} larvae in the region where its density is highest,
and continues its march rightward. However, on the left end of the
\emph{Torymus} peak, the density of the parasitoid is so low that
it is unable to control the pest. Therefore, the very small amount
of \emph{Dryocosmus} larvae originating from the eggs that were laid
on the extreme left of the \emph{Torymus} density peak, are left virtually
unaffected by the presence of the parasitoid (Figure \ref{fig:recolonization},
panel ``Year 8 - End of \emph{Torymus} season''). Thus, they are
able to develop into \emph{Dryocosmus} adults, that find, on their
left, a forest devoid of the parasitoid and ready to be recolonized
(Figure \ref{fig:recolonization}, panel ``Year 8 - End of \emph{Dryocosmus}
season''). In the next year the recolonized patch widens to the left,
and the density of the pest increases (Figure \ref{fig:recolonization},
``Year 9'' panels). In the following years (not pictured in Figure
\ref{fig:recolonization}), when the pest density has recovered to
sufficiently high density values, a second peak of the parasitoid
population splits from the first, sweeping leftward the recolonized
forest. Subsequently, the pest passes back through this second peak,
just as it did with the first. With the choice of parameters of the
run in Figure \ref{fig:recolonization}, the long term dynamics is
a never ending alternation of local extinctions and recolonizations.

The inability of the parasitoid to control the pest at low densities
of both species derives from the very low probability of finding egg
deposition sites when both host and parasitoid are rare. This is a
general characteristic of predator--prey systems and the ultimate
source of their cyclic behavior. In the case of the present model,
for spatially homogeneous solutions, a mathematical analysis of this
effect is given in §\ref{sub:Cycles-around-the-FP} (equation (\ref{eq:small_q_map})
and the following discussion). For solutions that have a dependence
on space, local population flows caused by diffusion become important,
and this means that regions where the pest had been eradicated and
thus the parasitoid has dropped to densities at which it is unable
to exert an effective control, may come again within reach of the
diffusing pest population, as we have illustrated discussing the Figure
\ref{fig:recolonization}.

In order to understand under which conditions the pest is able to
cross the parasitoid peaks and recolonize the forest, we have examined
a large sample of numerical solutions of the model equations, with
different parameters. The general pattern that emerges is the following:
if the speed of propagation of \emph{Torymus} peaks (as given by eq.
(\ref{eq:Torymus_front_speed})) is appreciably larger than that of
\emph{Dryocosmus} fronts (eq. (\ref{eq:cinipide_speed})), then the
pest will not be able to recolonize the forest. Conversely, if the
speed of \emph{Dryocosmus} fronts is sufficiently larger than that
of \emph{Torymus} peaks, then recolonization occurs. The precise boundary
between the two regimes is determined by the value of the threshold
of local extinction. 

\begin{figure}
\begin{centering}
\includegraphics[width=0.99\textwidth]{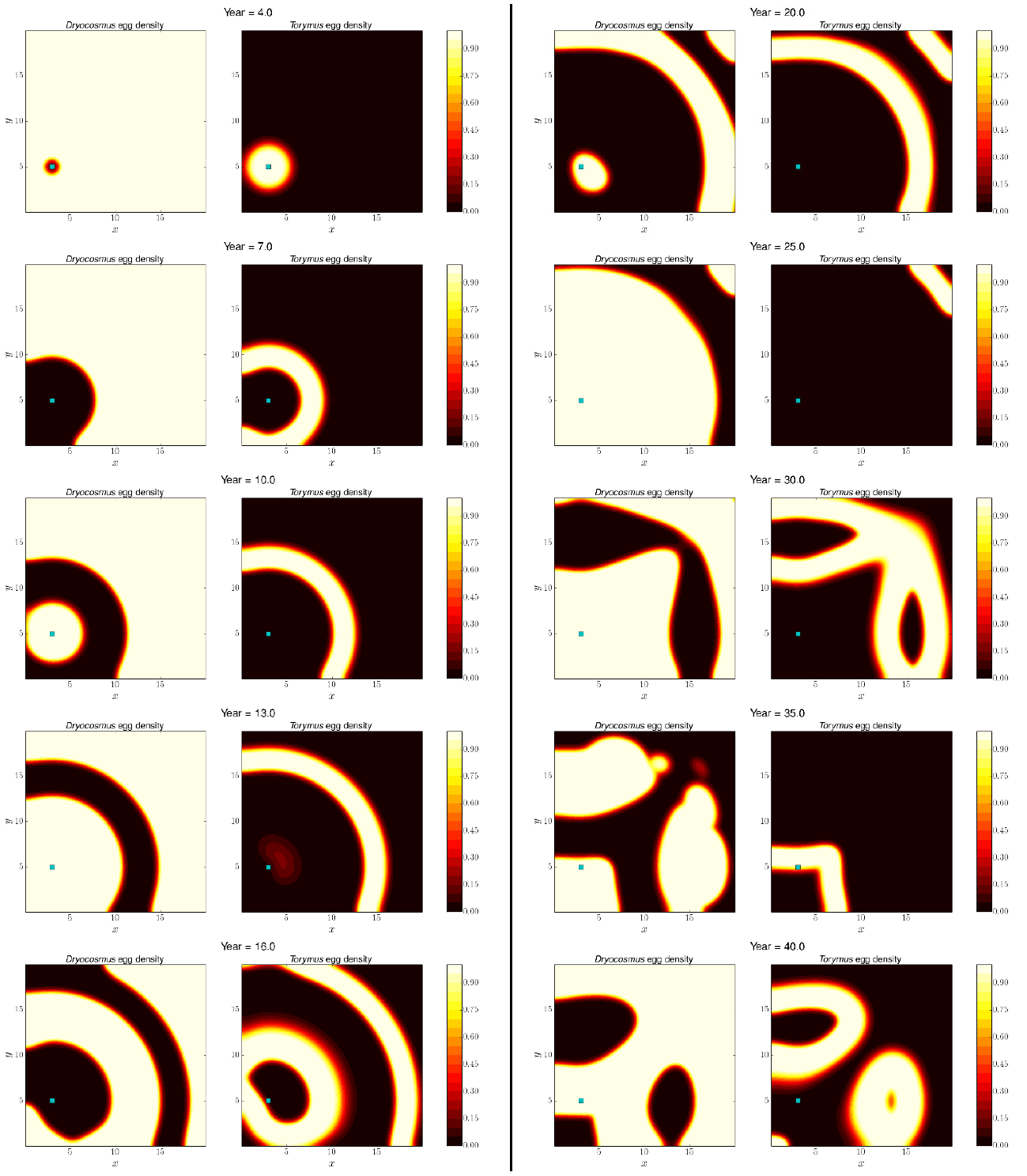}
\par\end{centering}

\caption{\label{fig:twoD}Time sequence of \emph{D. kuriphilus} (left panels)
and \emph{T. sinensis} (right panels) egg densities in a numerical
solution of the model equations with $\delta=0.2$ and the other parameters
as in Figure \ref{fig:recolonization}. The left column shows the
earlier years after the release of a small amount of \emph{T. sinensis}
in a small patch of a square forest saturated by \emph{D. kuriphilus}.
The right column shows the dynamics on a longer time scale. The marker
visible close to the lower left corner of all the panels is the release
site of \emph{T. sinensis}, and the place where the egg densities
shown in Figure \ref{fig:twoD_measurements_release_site} are measured.}
\end{figure}

\begin{figure}
\begin{centering}
\includegraphics[width=0.5\textwidth]{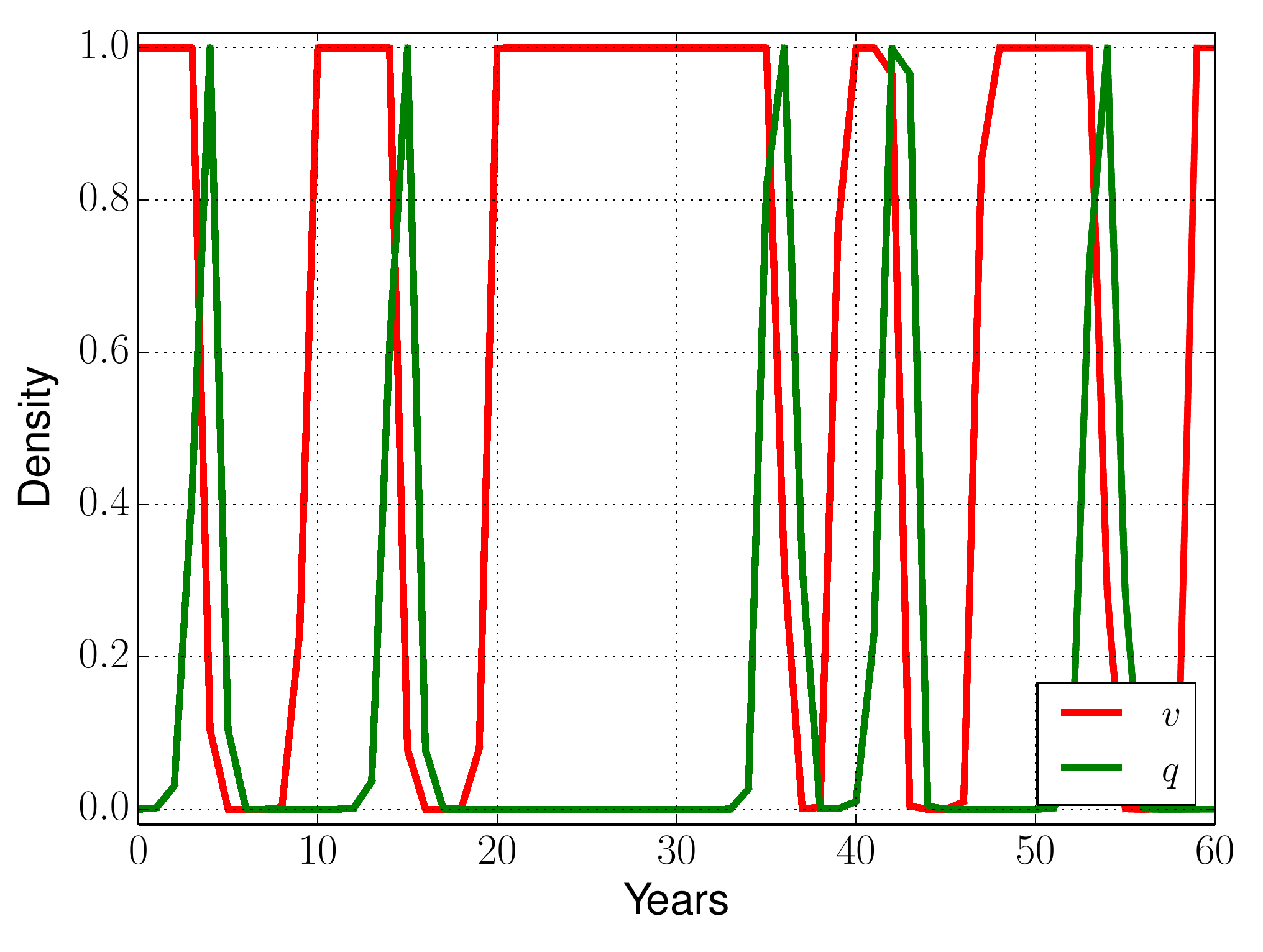}
\par\end{centering}

\caption{\label{fig:twoD_measurements_release_site}Densities of \emph{D. kuriphilus}
($v$) and \emph{T. sinensis} ($q$) eggs, at the end of each year,
measured at the \emph{T. sinensis} release site (shown in Figure \ref{fig:twoD})
in the numerical solution of the model equations shown in Figure \ref{fig:twoD}.}
\end{figure}

The dynamics of traveling fronts of \emph{Dryocosmus} and of sweeping
peaks of \emph{Torymus} is present also when the spatial domain is
two-dimensional. In this case, peaks and fronts may be verbally described
as waves propagating through the forest, as illustrated in the time
sequence of Figure \ref{fig:twoD}. The \emph{Torymus} is initially
released close to the lower--right corner of a square forest (the
size of $20\times20$ non--dimensional units corresponds in this numerical
solution to a physical size of approximately $110\times110$ km).
It spreads radially, leaving a roughly circular area of the forest
free of both the pest and the parasitoid, which is quickly recolonized
by the pest. The parasitoid population then splits in two parts: most
of it continues to propagate through the forest in an expanding arched
wave, and a small part returns close to the release site, hunting
the pest that has recolonized the release area, giving rise to a second
expanding arching wave (left column of Figure \ref{fig:twoD}). On
longer time scales, because of the interaction with the forest boundaries,
these waves assume irregular shapes and form a large variety of patterns
(right column of Figure \ref{fig:twoD}). The basic recolonization
mechanism, however, remains the same, and repeats endlessly. 

This means that the time scales of appearance and disappearance through
the years of both pest and parasitoid at any fixed place in the forest
are determined by the speed of propagation of the waves, and by the
size and shape of the forest itself. This is illustrated in Figure
\ref{fig:twoD_measurements_release_site}, showing the end--of--year
egg density of \emph{Dryocosmus} and \emph{Torymus} at the \emph{Torymus}
release site, in the numerical solution of Figure \ref{fig:twoD}.
In this case we have tuned the diffusivity ratio (namely, we used
$\delta=0.2$) explicitly to obtain cycles of pest and parasitoid
on a decadal time scale (roughly the same time scale as observed in
Japan). Note that, with these parameters, and neglecting the extinction
threshold, homogeneous solutions would give the cycles shown in Figure
\ref{fig:eta90_gamma45}, that have a much longer time scale. The
space--dependent solutions, instead, have the same time scale of the
homogeneous solutions of Figure \ref{fig:eta60_gamma20}, but with
much higher overwintering survival fractions.

\section{Discussion and conclusions\label{sec:Discussion-and-conclusions}}

In this paper we have developed a spatially explicit model that describes
the invasion of a chestnut forest by the gall wasp \emph{Dryocosmus
kuriphilus}, which acts as pest of the chestnut outside its native
China, as well as the effect of the parasitoid \emph{Torymus sinensis},
which is modeled as host--specific of the gall wasp and perfectly
synchronized to its life cycle. In the special case of a spatially
homogeneous distribution of both pest and parasitoid, the model can
be reduced to an iterated map. Otherwise it is a set of piecewise
time--continuous reaction--diffusion partial differential equations
which describe the spread and the competition for egg deposition sites
of the adults of both species. 

The primary aim of the model is that of elucidating the possibilities
of obtaining a biological control of the gall wasp, and understanding
possible causes of failure in obtaining control. In this respect the
crucial parameters are the overwintering survival fractions (the fraction
of laid eggs that successfully emerge the next year) and the diffusion
coefficients of the two species.

If the overwintering survival fractions are sufficiently far from
100\%, then both the spatially explicit model and its spatially--independent
counterpart show persistent oscillations in the density of both species,
reminiscent of the classic predator--prey models, having an amplitude
too small to be consistent with local extinction. This is in agreement
with the hypothesis of \citet{Murakami91}, that attributes the failure
of achieving biological control in Japan to the presence of non--specialist
parasitoid and hyperparasitoids species, which would cause a high
overwintering mortality. 

Our own observations strongly suggest that, in an European setting,
the overwintering survival fractions are, at least, 90\%. In that
case, in a spatially homogeneous situation, the model exhibits drops
of more than 15 orders of magnitude in the density of the pest, followed
by even larger drops in the density of the parasitoid. In practical
terms, drops of this magnitude can only be interpreted as signaling
the local extinction of the insect. Thus, in a spatially homogeneous
situation, and with parameters that we consider appropriate for the
European setting, the model predicts that the parasitoid would quickly
eradicate the pest.

However, ever since the seminal experiments of \citet{Huffaker58}
on mites, it is known that spatial inhomogeneities may delay or altogether
avoid phenomena of local extinction in predator--prey systems. In
particular, spatially explicit versions of the Nicholson--Bailey host--parasitoid
model show that the dispersal of the individuals spontaneously produces
the formation of complicated, time-varying, but persistent patterns.
In these cases, forcing a spatially--homogeneous environment (e.g.
by reducing the size of the domain below the intrinsic scale of the
patterns) often results in a rapid local extinction of both the host
and the parasitoid \citep{Hassel91}. 

With our model we find a similar outcome. If the speed of propagation
of the host population is faster than that of the parasitoid population
(those speeds are strongly dependent on the diffusion coefficients
of the two species) and the size of the domain is sufficiently large,
then the spatially explicit model never settles into a spatially--homogeneous
solution leading to extinction. Instead, the gall wasp recolonizes
the areas left empty after the passage of the parasitoid, in a never
ending sequence of crossing waves of population density, of which
Figure \ref{fig:twoD} shows an example. We should note that the imperfect
biological control achieved in Japan should probably be explained
in this way, because the attack rate of indigenous parasitoids was
later found to be no larger than 2\% \citep{EFSA10}, thus making
unlikely the hypothesis of low overwintering survival rates. 

On the other hand, when the population of the parasitoid propagates
sufficiently faster than that of the pest, then a single density wave
of the parasitoid sweeps the forest, killing the host, and leaving
neither species behind it. In this case the end result of the spatially
explicit and of the spatially homogeneous models is the same, and
they both suggest a complete eradication of the pest.

Quite remarkably, the literature does not appear to contain any quantitative
observation of the speed of propagation of \emph{Torymus sinensis}
when it is released in a forest fully invaded by \emph{Dryocosmus
kuriphilus}. In the United States, after the release there has been
a 30 years lapse with no follow-up observations \citep{CooperRieske07}.
In Japan the published data refer to insect densities at the release
site \citep{Moriya89,Murakami01,Moriya03} but give no information
on the spatial patterns of the insect population densities and their
changes in time. In Europe the releases of \emph{T. sinensis} have
begun only a few years ago, and we are not aware of surveys assessing
their spatial distribution in the following years. On the basis of
qualitative personal observations the authors suspect that \emph{T.
sinensis} actually spreads at a much lower rate than \emph{D. kuriphilus}.
If this were confirmed by quantitative measurements, then, according
to the model, we should expect that the release of \emph{T. sinensis}
at a single location within a large forest area would simply trigger
a train of density waves of both the pest and the parasitoid, that
would travel into the forest producing, at any fixed site, an alternating
presence and absence of the insects on decadal time scales. A satisfactory
control would then only be achieved by follow--up releases of\emph{
}the parasitoid, continuing for many years,\emph{ }at carefully chosen
sites, in order to suppress any returning wave of \emph{D. kuriphilus}
that could recolonize the forest left empty by the previous sweep
of \emph{T. sinensis}. This strategy, obviously, calls for a campaign
of accurate observations tracking the spatial distribution of both
species in the course of several years.

An important part of the model is the choice of describing the dispersion
of the individuals with diffusion operators. This is correct if the
overall motion of the individuals may be characterized by random flights,
and the length of the flights has a distribution with Gaussian or
shorter tails. Recently, it emerged that some animals appear to move
by performing so--called Lévy walks, that is, random trajectories
approximated by sequences of straight segments, where the probability
distribution of the lengths of each segment has long, algebraic tails,
and the variance of the distribution diverges. This movement strategy
should be advantageous when searching for randomly scattered resources,
in particular if they have a fractal distribution \citep{Viswanathan08}.
If \emph{D. kuriphilus} or \emph{T. sinensis} adopted this strategy,
then the model would have to be corrected with the use of fractional
diffusion operators, rather than ordinary ones. However, under the
Lévy walk hypothesis, a substantial fraction of the length traveled
by each insect during its lifetime would be traveled with just a few
straight legs. While straight flight is very likely to occur on scales
of the order of a meter, where visual and olfactory cues may reliably
direct the animal straight to its target, lacking published evidence,
it seems far--fetched to assume that the individuals of \emph{D. kuriphilus}
or \emph{T. sinensis} are consistently able to travel several hundred
of meters, or kilometers, along a straight line, within the canopy
of a forest, maintaining a precise bearing. We find further support
in our belief that \emph{D. kuriphilus} and \emph{T. sinensis} do
not perform Lévy walks from the mathematical result that equations
of the sort discussed in §\ref{sub:Space-dependent-dynamics}, but
with fractional diffusion, would generate traveling fronts that exponentially
accelerate, rather than maintain a constant speed \citep{Del Castillo 03}.
We are not aware of any report of acceleration of a gall wasp invasion
front. 

The fact that Lévy walks are unlikely to be the dispersion mechanism
of the pest and of the parasitoid, does not mean that other dispersion
processes, in addition to ordinary diffusion, should be ruled out.
The gall wasp spreading model of EFSA \citep{EFSA10} includes so--called
long--distance dispersal (LDD) events. There are essentially two main
causes for LDD events: transport due to antropic activities, and transport
with the wind of individuals that ventured above the forest canopy
(particularly on occasion of storms). Transport processes can easily
produce patchiness (see, for example, the case of zebra mussels carried
downstream a river: \citealp{Mari09}). In addition, in our case,
they are also completely random and unpredictable: while the outcome
of each LDD event could be forecast by a model, after its occurrence,
and provided the availability of sufficient observational data, the
occurrence of the event itself can not be forecast. Because our model
does not (yet) have the ambition of being an operational one, but
it is meant to uncover and elucidate some ecological processes in
the interaction between \emph{D. kuriphilus} and \emph{T. sinensis},
we have, for the moment, refrained from including LDDs into it.

However, we do not expect LDDs to change the overall picture that
has emerged about the likeliness of achieving biological control of
\emph{D. kuriphilus} with \emph{T. sinensis}. In fact, a random LDD
event involving \emph{D. kuriphilus} may have a good chance of carrying
the pest in a region where it is absent, thus creating a new hotspot
of infestation. On the other hand, a random LDD event involving \emph{T.
sinensis} can only contribute to the effectiveness of biological control
if the parasitoid lands in a region populated by the pest and devoid
of the parasitoid. If it lands in an area where the pest is absent,
the event has no effect. If it lands close to a parasitoid sweeping
wave, that region would have been swept in any case, and thus the
effect is also limited. Thus, it seems reasonable to assume that LDDs
do not improve the chances of achieving control, and, if anything,
they diminish them. Overall the message remains the same: biological
control of \emph{D. kuriphilus} with \emph{T. sinensis} may be a viable
option, but only if one is prepared to carefully track the distribution
of both species and to suppress any new hotspots (or recolonization
waves) with further releases of the parasitoid.

\section{Appendix\label{sec:Appendix}}

\subsection{\label{sub:Bounds}Approximations of the space-independent solution
of the gall wasp equations}

Let us define $y_{n}(t)=\log(1-v_{n}(t))$. Substituting in (\ref{eq:ODEs})
we obtain
\begin{equation}
\begin{cases}
\dot{u}_{n}(t)= & -\mu\left(1+e^{y_{n}(t)}\right)u_{n}(t)+v_{n-1}(1)\\
\dot{y}_{n}(t)= & -E_{D}\mu u_{n}(t)\\
u_{n}(0)\,= & 0\\
y_{n}(0)\,= & 0
\end{cases}\label{eq:ODEs_y}
\end{equation}
Observe that in (\ref{eq:ODEs_y}), because the egg density $v_{n}$
obeys $0\le v_{n}<1$, then $-\infty<y\le0$, and thus it is $1<1+e^{y_{n}(t)}\le2$.
This implies that
\begin{equation}
\frac{v_{n-1}(1)}{\mu}\left(1-e^{-\mu t}\right)>u_{n}(t)\ge\frac{v_{n-1}(1)}{2\mu}\left(1-e^{-2\mu t}\right)\label{eq:u-inequality}
\end{equation}
Using these inequalities in the second of (\ref{eq:ODEs_y}) and from
$v_{n}(t)=1-e^{y_{n}(t)}$, follow the inequalities
\begin{equation}
1-e^{-\frac{E_{D}}{\mu}\left(e^{-\mu t}+\mu t-1\right)v_{n-1}(1)}>v_{n}(t)>1-e^{-\frac{E_{D}}{4\mu}\left(e^{-2\mu t}+2\mu t-1\right)v_{n-1}(1)}.\label{eq:bounds}
\end{equation}
By evaluating the above expression at time $t=1$ one finds that the
year--over--year evolution of the end of season egg density $v_{n}(1)$
may be approximated by the map (\ref{eq:Skellam-model}) with the
constants (\ref{eq:Skellam_constant}).

Note that the approximation from below, obtained by choosing $k_{-}$
in (\ref{eq:Skellam_constant}), is very accurate if the density of
eggs laid in the previous year is low. In fact, by taking $v_{n-1}(1)$
arbitrarily close to zero it is possible, from (\ref{eq:u-inequality}),
to keep$u_{n}(t)$ as small as one wishes, for all $t\in[0,1]$, and,
from the second of (\ref{eq:ODEs_y}) also $y_{n}(t)$ may be kept
as close to zero as one wishes, for all $t\in[0,1]$. Therefore, the
quantity $1+e^{y_{n}(t)}$ may be kept arbitrarily close to $2$,
which is the value used by the approximation from below. Conversely,
if $v_{n-1}(1)$ is close to one, and the product $E_{D}\mu$ is much
larger than one, then $1+e^{y_{n}(t)}$ will rapidly approach the
value $1$. Therefore, we expect the approximation from above to be
more accurate at high densities of eggs laid in the previous year.

\subsection{\label{sub:Exact_solutions_Torymus}Exact space-independent solution
of the equations for \emph{T. sinensis}}

By imposing no-flux boundary conditions on $p_{n}$, assuming that
$\nabla v_{n-1}=\nabla q_{n-1}=0$, the equations (\ref{eq:Torymus_equations_nondim})
and the conditions (\ref{eq:Torymus_initial_conditions-nondim}) become
\begin{equation}
\begin{cases}
{\displaystyle \dot{p}_{n}(t)\,=} & {\displaystyle -\tau^{-1}\left(v_{n-1}(1)-q_{n}(t)\right)p_{n}(t)}\\
{\displaystyle \dot{q}_{n}(t)\,=} & E_{T}\tau^{-1}\left(v_{n-1}(1)-q_{n}(t)\right)p_{n}(t)\\
p_{n}(0)\,= & q_{n-1}(\eta\tau)\\
q_{n}(0)\,= & 0
\end{cases}.\label{eq:Torymus_ode}
\end{equation}
Dividing the first by the second we have
\[
\dot{p}_{n}=-\frac{1}{E_{T}}\dot{q}_{n},
\]
and thus, by integration and using the initial conditions, we find
\[
p_{n}(q_{n}(t))=-\frac{q_{n}(t)}{E_{T}}+q_{n-1}(\eta\tau).
\]
Substituting this expression in the second of the equations (\ref{eq:Torymus_ode})
we obtain a first-order, autonomous equation for $q_{n}$ which yields
the following solution
\[
\begin{cases}
\begin{cases}
p_{n}(t) & =\frac{\bar{q}_{n-1}\left(E_{T}\bar{q}_{n-1}-\bar{v}_{n-1}\right)\exp\left(\frac{t}{\tau}\left(E_{T}\bar{q}_{n-1}-\bar{v}_{n-1}\right)\right)}{\bar{q}_{n-1}E_{T}\exp\left(\frac{t}{\tau}\left(E_{T}\bar{q}_{n-1}-\bar{v}_{n-1}\right)\right)-v_{n-1}^{*}}\\
q_{n}(t) & =\frac{\bar{v}_{n-1}\bar{q}_{n-1}E_{T}\left(1-\exp\left(\frac{t}{\tau}\left(E_{T}\bar{q}_{n-1}-\bar{v}_{n-1}\right)\right)\right)}{\bar{v}_{n-1}-\bar{q}_{n-1}E_{T}\exp\left(\frac{t}{\tau}\left(E_{T}\bar{q}_{n-1}-\bar{v}_{n-1}\right)\right)}
\end{cases}, & E_{T}\bar{q}_{n-1}\neq\bar{v}_{n-1}\\
\begin{cases}
p_{n}(t) & =\frac{\bar{v}_{n-1}\tau}{E_{T}\left(\bar{v}_{n-1}t+\tau\right)}\\
q_{n}(t) & =\frac{\bar{v}_{n-1}^{2}t}{\bar{v}_{n-1}t+\tau}
\end{cases}, & E_{T}\bar{q}_{n-1}=\bar{v}_{n-1}
\end{cases}
\]
where, for brevity, we have defined the shorthands $\bar{q}_{n-1}=q_{n-1}(\eta\tau)$,
$\bar{v}_{n-1}=v_{n-1}(1)$. It can be verified, by expanding the
exponentials in power series, that the above solution is a smooth
function of the quantity $E_{T}\bar{q}_{n-1}-\bar{v}_{n-1}$.

\subsection{Mathematical properties of the space--independent map\label{sub:Mathematical-properties-of-map}}

\subsubsection{\label{sub:Global-boundedness}Boundedness of the global dynamics}

The map (\ref{eq:complete_map}), formally, does not allow for the
extinction of either species. Specifically, the map has the property
that, if $E_{T},\eta,k>0$ and $0<v_{n},q_{n}$ then $0<v_{n+1},q_{n+2}<1$. 

This assert becomes apparent by rewriting the map in the following
form
\begin{equation}
\begin{cases}
q_{n+1}= & \begin{cases}
\left({\displaystyle \frac{1-e^{\eta\left(E_{T}q_{n}-v_{n}\right)}}{\frac{v_{n}}{E_{T}q_{n}}-e^{\eta\left(E_{T}q_{n}-v_{n}\right)}}}\right)v_{n}, & E_{T}q_{n}\neq v_{n}\\
\left(\frac{v_{n}}{\eta^{-1}+v_{n}}\right)v_{n}, & E_{T}q_{n}=v_{n}
\end{cases}\\
v_{n+1}= & 1-e^{-k\left(v_{n}-q_{n+1}\right)}
\end{cases}\label{eq:mappa_cinipide_torymus_riscritta}
\end{equation}
It is straightforward to verify that 
\[
0<\left[1-\exp\left(\eta\left(E_{T}q_{n}-v_{n}\right)\right)\right]\left[v_{n}E_{T}^{-1}q_{n}^{-1}-\exp\left(\eta\left(E_{T}q_{n}-v_{n}\right)\right)\right]^{-1}<1
\]
 both if $E_{T}q_{n}<v_{n}$ and if $E_{T}q_{n}>v_{n}$. Obviously,
it it also 
\[
0<v_{n}\left[\eta^{-1}+v_{n}\right]^{-1}<1,
\]
which is relevant in the case $E_{T}q_{n}=v_{n}$. Therefore, from
the equation for $q_{n+1}$ in (\ref{eq:mappa_cinipide_torymus_riscritta})
we have $0<q_{n+1}<v_{n}$. Using this inequality in the equation
for $v_{n+1}$ in (\ref{eq:mappa_cinipide_torymus_riscritta}) we
have $0<v_{n+1}<1$, and, therefore $0<q_{n+2}<v_{n+1}<1$.

\subsubsection{Nullclines and the coexistence fixed point\label{sub:Nullclines}}

We defined $v-$nullcline as the set of pairs $(v_{n},q_{n+1})$ such
that $v_{n+1}=v_{n}$. From the second equation in (\ref{eq:complete_map}),
we find that the $v-$nullcline has the following explicit expression
\begin{equation}
q_{n+1}(v_{n})=v_{n}+\frac{1}{k}\log\left(1-v_{n}\right)\label{eq:v-nullcline}
\end{equation}
whose graph is the red line in the right panel of Figures \ref{fig:eta35_gamma15},
\ref{fig:eta60_gamma20}, \ref{fig:eta90_gamma45}. A simple calculation
shows that if $(v_{n},q_{n+1})$ is above the $v-$nullcline, then
$v_{n+1}<v_{n}$, and if it is below, then $v_{n+1}>v_{n}$. If $k>1$
then the $v-$nullcline has a maximum at 
\begin{equation}
v_{mx}=\frac{k-1}{k},\quad q_{mx}=\frac{k-1-\log(k)}{k}.\label{eq:v-nullcine_max}
\end{equation}
It also has a zero at $v_{n}=0$, and at a value larger than $v_{mx}$
and smaller than 1, which does not have a simple explicit expression,
and corresponds to the non-zero fixed point of Skellam's map (\ref{eq:Skellam-model}). 

Analogously we defined the $q-$nullcline as the set of pairs $(v_{n},q_{n})$
such that $q_{n+1}=q_{n}$. The $q-$nullcline has an obvious branch
which is $q_{n}=0$. From the first equation in (\ref{eq:complete_map}),
if $v_{n},q_{n}\neq0$ and $E_{T}q_{n}\neq v_{n}$, we have the following
implicit definition of the $q-$nullcline
\begin{equation}
\frac{v_{n}-q_{n}}{v_{n}}e^{\eta(E_{T}q_{n}-v_{n})}=\frac{E_{T}-1}{E_{T}}.\label{eq:q-nullcline_implicit}
\end{equation}
In the case $E_{T}q_{n}=v_{n}>0$, it is straightforward to verify
from (\ref{eq:complete_map}) that only the point $\left(v_{n},q_{n}\right)=\left(\eta^{-1}(E_{T}-1)^{-1},\,\eta^{-1}(E_{T}-1)^{-1}E_{T}^{-1}\right)$
belongs to the $q-$nullcline, shown as the green curve in the right
panel of Figures \ref{fig:eta35_gamma15}, \ref{fig:eta60_gamma20},
\ref{fig:eta90_gamma45}. If $q_{n}\to0$, the $q-$nullcline tends
to the value 
\begin{equation}
v_{z}=\frac{1}{\eta}\log\left(\frac{E_{T}}{E_{T}-1}\right).\label{eq:zero_of_q-nullcline}
\end{equation}
Note that, if $E_{T}>1$ then the right--hand side of (\ref{eq:q-nullcline_implicit})
is larger than 0 and smaller than 1. Thus, for fixed $\eta$ and $v_{n}\neq v_{z}$,
if $E_{T}\to\infty$ either there is no solution to (\ref{eq:q-nullcline_implicit}),
or $q_{n}\to v_{n}$ from below. This observation suggests that for
realistic values of the parameters (that is $\eta$ not much smaller
than 1 and $E_{T}$ quite larger than 10), taking $q_{n}\approx v_{n}$
for $v_{n}>v_{z}$ should give a reasonably good approximation of
the $q-$nullcline.

With respect to the new variable $z=q_{n}/v_{n}$, the implicit expression
(\ref{eq:q-nullcline_implicit}) may be made explicit, and one finds
\begin{equation}
v_{n}(z)=\frac{1}{\eta\left(E_{T}z-1\right)}\log\left(\frac{E_{T}-1}{E_{T}-E_{T}z}\right)\label{eq:vn(z)}
\end{equation}
Note that $0<z<1$ because $q_{n}=q_{n+1}<v_{n}$. It can be checked
that this is a strictly growing function of $z$, which is smooth
even at $z=E_{T}^{-1}$. (In order to verify the positive sign of
the derivative the identity $\log(x)\le x-1$ can be useful.) Thus
the minimum of this function is attained in the limit $z\to0$, where
$v_{n}\to v_{z}$: for $v_{n}<v_{z}$ the equation (\ref{eq:q-nullcline_implicit})
has no solution. We also observe that, because $v_{n}^{\prime}(z)>0$,
to each value of $z$ corresponds a unique value of $q_{n}(z)=zv_{n}(z)$.
Therefore, calling $\zeta$ the inverse function of (\ref{eq:vn(z)}),
we have that the equation (\ref{eq:q-nullcline_implicit}) implicitly
defines a unique continuous function $q_{n}(v_{n})=v_{n}\zeta(v_{n})$
of $v_{n}$, that we shall call the non-zero branch of the $q-$nullcline,
and that $q_{n}^{\prime}(v_{n})>0$, as depicted by the green line
in the right panel of Figures \ref{fig:eta35_gamma15}, \ref{fig:eta60_gamma20},
\ref{fig:eta90_gamma45}.

From the first equation in (\ref{eq:small_q_map}) below, (see also
the surrounding discussion) it is clear that, for states not belonging
to the $q-$nullcline having arbitrarily small $q_{n}$, if $v_{n}<v_{z}$
then $q_{n+1}<q_{n}$, and if $v_{n}>v_{z}$ then $q_{n+1}>q_{n}$.
Thus, since the non-zero branch of the $q-$nullcline is unique, by
the theorem of the persistence of sign, if a state $(v_{n},q_{n})$
lies on the left of the $q-$nullcline, then $q_{n+1}<q_{n}$; if
it lies on the right, then $q_{n+1}>q_{n}$. 

Fixed points are the intersection of the $v-$nullcline and of the
$q-$nullcline. There is always the fixed point $(v_{n},q_{n})=(0,0)$.
If $k>1$ then there is also the fixed point $(v_{n},q_{n})=(v^{*},0)$
where $v^{*}$ is the non-zero fixed point of Skellam's map (\ref{eq:Skellam-model}). 

If $k>1$ and $v_{z}<v^{*}$, then the non--zero branch of the $q-$nullcline
(which is a growing function of $v_{n}$) must cross at at least one
point the $v-$nullcline (which is positive and has a zero at $v_{n}=0$
and a zero at $v_{n}=v^{*}$). We call this is a \emph{coexistence}
fixed point, because both $v_{n}$ and $q_{n}$ are larger than 0.
Note that, except for unrealistically low values of $k$, the non-zero
fixed point of Skellam's map is very close to one. Thus, an approximate
criterion for the existence of a coexistence fixed point is $v_{z}<1$.
Using the expression (\ref{eq:zero_of_q-nullcline}) and the definition
of $E_{T}$ (see Table \ref{tab:Rescaled-parameters}), setting $v_{z}=1$
one obtains the approximate threshold (\ref{eq:gamma-treshold}).

We have ample numerical evidence, corroborated by asymptotic results,
that there is only one coexistence fixed point, although we cannot
exclude that for some finely--tuned value of the parameters more than
one coexistence fixed point could exist. 

In the realistic range of parameters, a very rough approximation of
the position of the coexistence fixed point may be obtained by approximating
the $v-$nullcline as
\[
q_{n+1}(v_{n})\approx\left(1-\frac{1}{k}\right)v_{n}
\]
and the $q-$nullcline as the straight line connecting the points
\[
\left(v_{n},q_{n}\right)=\left(\eta^{-1}(E_{T}-1)^{-1},\,\eta^{-1}(E_{T}-1)^{-1}E_{T}^{-1}\right)\quad\mathrm{and\quad}\left(v_{n},q_{n}\right)=\left(1,\,1\right).
\]
 Looking for the intersection of these straight lines we find
\begin{equation}
\begin{cases}
v_{c}= & \frac{k\left(E_{T}-1\right)}{E_{T}\left[\eta\left(E_{T}-1\right)-1\right]+k\left(E_{T}-1\right)},\\
q_{c}= & \frac{\left(k-1\right)\left(E_{T}-1\right)}{E_{T}\left[\eta\left(E_{T}-1\right)-1\right]+k\left(E_{T}-1\right)}.
\end{cases}\label{eq:approx_coexistence_FP}
\end{equation}
More accurate approximations of the $q-$nullcline (and thus of the
coexistence fixed point) can be worked out by evaluating (\ref{eq:vn(z)})
at the values $z_{m}$ such that 
\[
\frac{E_{T}-1}{E_{T}-E_{T}z_{m}}=e^{\eta E_{T}/m}
\]
for distinct values of the arbitrary parameter $m$. This yields explicit
expressions of points $\left(v_{n}(z_{m}),q_{n}(z_{m})\right)$ lying
on the $q-$nullcline among which it is possible to interpolate with
any standard method.

\subsubsection{Cycles around the fixed point\label{sub:Cycles-around-the-FP}}

The cyclic dynamics generated by the map (\ref{eq:complete_map})
may be qualitatively understood through the following argument. For
small $q_{n}$, the map (\ref{eq:complete_map}) becomes, at leading
order
\begin{equation}
\begin{cases}
q_{n+1}= & \left(1-e^{-\eta v_{n}}\right)E_{T}q_{n}+O(q_{n}^{2})\\
v_{n+1}= & \left(1-e^{-kv_{n}}\right)\hphantom{E_{T}q_{n}}+O(v_{n}q_{n})
\end{cases}.\label{eq:small_q_map}
\end{equation}
Let us assume that initially $v_{n}$ is also very small. Thus the
egg density of \emph{T. sinensis} decreases from one year to the next
as long as $\left(1-e^{-\eta v_{n}}\right)<E_{T}^{-1}$, that is,
until $v_{n}<v_{z}$ (see eq. \ref{eq:zero_of_q-nullcline}). However,
\emph{D. kuriphilus} is at leading order decoupled from its parasitoid,
and its egg density obeys Skellam's map (\ref{eq:Skellam-model}).
Therefore, assuming $k>1$, $v_{n}$ will grow with $n$ until it
approaches the non-zero fixed point of Skellam's map, which, for realistic
values of $k$, has a numerical value very close to 1. At this point
the egg density of \emph{T. sinensis} will be growing in time, but
it may require several years before reaching an $O(1)$ magnitude.
Thus, starting from very small values of $v_{n}$ and $q_{n}$, we
have that the sequence of states, seen in a diagram $q_{n}$ vs $v_{n}$,
as in the right panel of Figure \ref{fig:eta90_gamma45}, first moves
horizontally ($v_{n}$ growing, $q_{n}$ very close to 0) and then
vertically ($v_{n}$ very close to 1, $q_{n}$ growing). When $q_{n}$
reaches $O(1)$ the approximation (\ref{eq:small_q_map}) no longer
applies, and it is convenient to rewrite the map (\ref{eq:complete_map})
as (\ref{eq:mappa_cinipide_torymus_riscritta}), and then (assuming
$E_{T}q_{n}\neq v_{n}$) as 
\begin{equation}
\begin{cases}
q_{n+1}= & \left({\displaystyle \frac{e^{-\eta\left(E_{T}q_{n}-v_{n}\right)}-1}{\frac{v_{n}}{E_{T}q_{n}}e^{-\eta\left(E_{T}q_{n}-v_{n}\right)}-1}}\right)v_{n}\\
v_{n+1}= & 1-e^{-k\left(v_{n}-q_{n+1}\right)}
\end{cases}\label{eq:mappa_cinipide_torymus_riscritta-2}
\end{equation}
With $q_{n}=O(1)$ for large (realistic) values of $E_{T}$ the parenthesis
appearing in the first equation approaches 1. Thus we have $q_{n+1}\approx v_{n}$,
which leads to a cancellation in the exponent appearing in the second
equation, causing a sharp drop in the value of $v_{n+1}$ with respect
to $v_{n}$. Thus the system jumps from a state ($v_{n}\approx1$,
$q_{n}=O(1)$) close to right edge of Figure \ref{fig:eta90_gamma45}
(right panel) to a state close to its upper edge ($v_{n+1}\approx O(1)$,
$q_{n+1}\approx1$) or, more often, depending on the exact value of
$q_{n}$, to a state close to its upper-left corner ($v_{n+1}\ll1$,
$q_{n+1}\approx1$). The next year, since $q_{n+1}\approx1$, the
cancellation occurs again, and the further drop in the value of \emph{D.
kuripilus}' egg density is as large as ten orders of magnitude, with
the parameters of Figure \ref{fig:eta90_gamma45}. Thus, in the turn
of just two years, \emph{T. sinensis} wipes out almost all the population
of \emph{D. kuripilus}, and, consequently, its own, because of the
constraint $q_{n+1}<v_{n}$. Then the cycle starts again. 

Note that the cycles need not be exactly periodic. In fact, the intervals
of exponential growth of $q_{n}$ and the subsequent cancellation
events could even produce a chaotic dynamics (however we did not investigate
this issue). More importantly, small differences in the value of $q_{n}$
before the cancellation events can make a large difference in the
number of orders of magnitude lost after the events, and thus in the
number of years needed to re--grow up to $O(1)$.


\begin{thebibliography}{del-Castillo-Negrete et al.(2003)}
\bibitem[Abe et al.(2007)]{Abe07}Abe Y., Melika G., Stone G. N. (2007)
The diversity and phylogeography of cynipid gallwasps (Hymenoptera,
Cynipidae) of the Eastern Palearctic and their associated communities.
Oriental Insects 41:169--212.

\bibitem[Alma et al.(2014)]{Alma14}Alma A., Ferracini C., Sartor
C., Ferrari E., Botta R. (2014) Il cinipide orientale del castagno:
lotta biologica e sensibilità varietale. Italus Hortus 21(3):15--29.

\bibitem[Balkowsky and Shraiman(2002)]{Balkovsky&Shraiman}Balkovsky
E., Shraiman B. I. (2002) Olfactory search at high Reynolds number.
PNAS 99(20):12589--12593.

\bibitem[Battisti et al.(2014)]{Battisti14}Battisti A., Benvegnù
I., Colombari F., Haack R. A. (2014) Invasion by the chestnut gall
wasp in Italy causes significant yield loss in \emph{Castanea sativa}
nut production. Agric. Forest. Entomol. 16(1): 75--79.

\bibitem[Borowiec et al.(2014)]{Borowiec14}Borowiec N., Thaon M.,
Brancaccio L., Warot S., Vercken E., Fauvergue X., Ris N., Malausa
J. C. (2014) Classical biological control against the chestnut gall
wasp \emph{Dryocosmus kuriphilus} (Hymenoptera, Cynipidae) in France.
Plant Prot. Q. 29(1):7--10. 

\bibitem[Bosio et al.(2013)]{Bosio13}Bosio G., Armando M., Moriya
S. (2013). Verso il controllo biologico del cinipide del castagno.
Informatore Agrario 4(14):60--64.

\bibitem[Bounous(2014)]{Bounous14}Bounous G. (2014) Il castagno:
risorsa multifunzionale in Italia e nel mondo. Edagricole, Bologna.

\bibitem[Br\"annstr\"om and Sumpter(2005)]{Brannstrom05}Br\"annstr\"om
\r{A}., Sumpter D. J. T. (2005). The role of competition and clustering
in population dynamics. Proc. R. Soc. B 272: 2065--2072.

\bibitem[Brussino et al.(2002)]{Brussino02}Brussino G., Bosio G.,
Baudino M., Giordano R., Ramello F., Melika G., (2002) Pericoloso
insetto esotico per il castagno europeo. Informatore Agrario, 58(37):59--61. 

\bibitem[Cooper and Rieske(2007)]{CooperRieske07}Cooper W. R., Rieske
L. K. (2007) Community associates of an exotic gallmaker, \emph{Dryocosmus
kuriphilus} (Hymenoptera: Cynipidae), in Eastern North America. Ann.
Entomol. Soc. Am. 100(2):236-244.

\bibitem[Cooper and Rieske(2010)]{CooperRieske10}Cooper W. R., Rieske
L. K. (2010). Gall Structure Affects Ecological Associations of \emph{Dryocosmus
kuriphilus} (Hymenoptera: Cynipidae). Environ. Entomol. 39(3):787--797.

\bibitem[Cho and Lee(1963)]{Cho&Lee63}Cho D. Y., Lee S. O. (1963)
Ecological studies on the chestnut gall wasp, \emph{Dryocosmus kuriphilus}
Yasumatsu, and observations on the damages of the chestnut trees by
its insect. Kor. J. Plant Protect. 2:47--54.

\bibitem[del-Castillo-Negrete et al.(2003)]{Del Castillo 03}del-Castillo-Negrete
D., Carreras B. A., Lynch V. E. (2003) Front dynamics in reaction-diffusion
systems with Levy flights: a fractional diffusion approach. Physical
Review Letters 91(1): 018302.

\bibitem[EFSA(2010)]{EFSA10}EFSA Panel on Plant Health (PLH) (2010)
Risk assessment of the oriental chestnut gall wasp, \emph{Dryocosmus
kuriphilus} for the EU territory on request from the European Commission.
EFSA J 8:1619

\bibitem[EPPO(2005)]{Eppo05}EPPO (2005) Data sheets on quarantine
pests-\emph{Dryocosmus kuriphilus}. EPPO Bull 35:422--424.

\bibitem[EPPO(2013)]{Eppo13}EPPO (2013) First report of \emph{Dryocosmus
kuriphilus} in Austria (2013/140, First report of \emph{Dryocosmus
kuriphilus} in Germany 2013/141, \emph{Dryocosmus kuriphilus} found
in Hungary 2013/142. Eppo Reporting service, No. 7, pp. 3--4.

\bibitem[EPPO(2015a)]{Eppo15}EPPO (2015a) First report of \emph{Dryocosmus
kuriphilus} in the United Kingdom. Eppo Reporting service, No. 6,
pp. 2.

\bibitem[EPPO(2015b)]{Eppo15b}EPPO (2015b) Dryocosmus kuriphilus
found again in the Netherlands 2015/128. Eppo Reporting Service no.7
p. 3.

\bibitem[Ferracini et al.(2015a)]{Ferracini15a}Ferracini C., Ferrari
E., Saladini M. A., Pontini M., Corradetti M., Alma A. (2015a) Non-target
host risk assessment for the parasitoid \emph{Torymus sinensis}. BioControl,
in press. DOI:10.1007/s10526-015-9676-1. 

\bibitem[Ferracini et al.(2015b)]{Ferracini15b}Ferracini C., Gonella
E., Ferrari E., Saladini M. A., Picciau L., Tota F., Pontini M., Alma
A. (2015b) Novel insight in the life cycle of \emph{Torymus sinensis},
biocontrol agent of the chestnut gall wasp. BioControl, 60:583--594.

\bibitem[Gardiner(2004)]{Gardiner04}Gardiner C. W.. (2004) Handbook
of Stochastic Methods. III edition. Springer-Verlag Berlin Heidelberg.

\bibitem[Germinara et al.(2011)]{Germinara11}Germinara G. S., De
Cristofaro A., Rotundo G. (2011) Chemical Cues for Host Location by
the Chestnut Gall Wasp, \emph{Dryocosmus kuriphilus}. J. Chem. Ecol.
37:49--56.

\bibitem[Gibbs et al.(2011)]{Gibbs11}Gibbs M., Schönrogge K., Alma
A., Melika G., Quacchia A., Stone G. N., Aebi A. (2011) \emph{Torymus
sinensis}: a viable management option for the biological control of
\emph{Dryocosmus kuriphilus} in Europe? BioControl 56:527--538.

\bibitem[Graziosi and Santi(2008)]{Graziosi08}Graziosi I., Santi
F. (2008) Chestnut gall wasp (\emph{Dryocosmus kuriphilus}): spreading
in Italy and new records in Bologna province. Bull. Insectol. 61(2):343--348.

\bibitem[Graziosi and Rieske(2013)]{Graziosi13}Graziosi I., Rieske
L. K. (2013) Response of \emph{Torymus sinensis}, a parasitoid of
the gallforming \emph{Dryocosmus kuriphilus}, to olfactory and visual
cues. Biological Control 67:137--142.

\bibitem[Graziosi and Rieske(2014)]{Graziosi14}Graziosi I., Rieske
L. K. (2014) Potential fecundity of a highly invasive gall maker,
\emph{Dryocosmus kuriphilus} (Hymenoptera: Cynipidae). Environmental
Entomology 43(4):1053--1058. 

\bibitem[Hassel et al.(1991)]{Hassel91}Hassell M., P., Comins H.
N., May R. M. (1991) Spatial structure and chaos in insect population
dynamics. Nature 353(6341):255--258.

\bibitem[Henneman et al.(2002)]{Henneman02}Henneman M. L., Dyreson
E. G., Takabayashi J., Raguso R. A. (2002) Response to walnut olfactory
and visual cues by the parasitc wasp \emph{Diachasmimorpha juglandis}.
J. Chem. Ecol. 28(11):2221--2244.

\bibitem[Huber and Read(2012)]{Huber12}Huber J. T., Read J. (2012)
First record of the oriental chestnut gall wasp, \emph{Dryocosmus
kuriphilus} Yasumatsu (Hymenoptera: Cynipidae), Canada. J. Entomol.
Soc. Ont. 143:125--128. 

\bibitem[Huffaker(1958)]{Huffaker58}Huffaker C. B. (1958) Experimental
Studies on Predation: Dispersion Factors and Predator--Prey Oscillations.
Hilgardia: A Journal of Agricultural Science 27:795--834.

\bibitem[Johansson et al.(2013)]{Mpmath}Johansson F. et al. (2013)
Mpmath: a Python library for arbitrary-precision floating-point arithmetic
(version 0.18). \url{http://mpmath.org/}

\bibitem[Kamijo(1982)]{Kamijo82}Kamijo K. (1982) Two new species
of \emph{Torymus} (Hymenoptera, Torymidae) reared from \emph{Dryocosmus
kuriphilus} (Hymenoptera, Cynipidae) in China and Korea. Kontyû 50:505--510.

\bibitem[Kato and Hijii(1997)]{Kato97}Kato K, Hijii N (1997) Effects
of gall formation by \emph{Dryocosmus kuriphilus} Yasumatsu (Hymenoptera:
Cynipidae) on the growth of chestnut trees. J. Appl. Entomol. 121:9--15.

\bibitem[Mari et al.(2009)]{Mari09}Mari L., Casagrandi R., Pisani
M. T., Pucci E., Gatto, M. (2009). When will the zebra mussel reach
Florence? A model for the spread of Dreissena polymorpha in the Arno
water system (Italy). Ecohydrology, 2(4):428--439.

\bibitem[Matoševi\'{c} et al.(2014)]{Mato=000161evi=00010714} Matoševi\'{c}
D., Quacchia A., Kriston E., Melika G., (2014). Biological control
of the invasive \emph{Dryocosmus kuriphilus} (Hymenoptera: Cynipidae)
- an overview and the first trials in Croatia. South-east European
Forestry, 5(1): 3--12. 

\bibitem[May and McLean(2007)]{TheoreticalEcology2007}May R. M.,
McLean A. R., (2007). Theoretical ecology: principles and applications.
Oxford University Press, III ed, Oxford (UK).

\bibitem[Moriya et al.(1989)]{Moriya89}Moriya S., Inoue K., Ôtake
A., Shiga M., Mabuchi M. (1989) Decline of the chestnut gall wasp
population, \emph{Dryocosmus kuriphilus} Yasumatsu (Hymenoptera: Cynipidae)
after the establishment of \emph{Torymus sinensis} Kamijo (Hymenoptera:
Torymidae). Appl Entomol Zool 24:231--233.

\bibitem[Moriya et al.(2003)]{Moriya03}Moriya S., Shiga M., Adachi
I. (2003) Classical biological control of the chestnut gall wasp in
Japan. In: Van Driesche R. G. (ed) Proceedings of the 1st International
Symposium on Biological Control of Arthropods. USDA Forest Service,
Washington.

\bibitem[Murakami et al.(1977)]{Murakami77}Murakami Y., Umeya K.,
Oho N. (1977) A preliminary introduction and released of a parasitoid
(Chalcidoidea, Torymidae) of the chestnut gall wasp \emph{Dryocosmus
kuriphilus} Yasumatsu. Jpn J Appl Entomol Zool 21:197--203.

\bibitem[Murakami et al.(1980)]{Murakami80}Murakami Y., Ao H. B.,
Chang C. H. (1980) Natural enemies of the chestnut gall wasp in Hopei
Province, China (Hymenoptera: Chalcidoidea). Appl. Entomol. Zool.
15:184--186.

\bibitem[Murakami(1981)]{Murakami81}Murakami Y. (1981) The parasitoids
of \emph{Dryocosmus kuriphilus} Yasumatsu (Hymenoptera: Cynipidae)
in Japan and the introduction of a promising natural enemy from China
(Hymenoptera: Chalcidoidea). J. Fac. Agric. Kyushu Univ. 25:167--174.

\bibitem[Murakami and Gyoutoku(1991)]{Murakami91}Murakami Y. and
Gyoutoku Y. (1991) Colonization of the imported \emph{Torymus (Syntomaspis)
sinensis} Kamijo (Hymenoptera: Torymidae) parasitic on the chestnut
gall wasp (Hymenoptera: cynipidae). (5) Mortality of \emph{Torymus}
spp. by native facultative hyperparasitoids. Proceedings of the Association
for Plant Protection of Kyushu 37: 194--197.

\bibitem[Murakami and Gyoutoku(1995)]{Murakami95}Murakami Y., Gyoutoku
Y. (1995) A delayed increase in the population of an imported parasitoid,
\emph{Torymus} (\emph{Syntomaspis}) \emph{sinensis} (Hymenoptera:
Torymidae) in Kumamoto, Southwestern Japan. Appl Entomol Zool 30:215--224.

\bibitem[Murakami et al.(2001)]{Murakami01}Murakami Y., Toda S.,
Gyoutoku Y. (2001) Colonization of imported \emph{Torymus} (\emph{Syntomaspis})
\emph{sinensis} Kamijo (Hymenoptera: Torymidae) parasitic on the chestnut
gall wasp (Hymenoptera: Cynipidae). Success in the eighteenth year
after release in Kumamoto. Proc. Assoc. Pl. Prot. Kyushu 47:132--134.

\bibitem[Murray(2007)]{Murray07}Murray J. D. (2007) Mathematical
Biology I. An Introduction. Springer, Berlin, 3rd ed.

\bibitem[NPPO(2013)]{NPPO13}NPPO The Netherlands (2013) Follow-up
pest report \emph{Dryocosmus kuriphilus}. Confirmation of eradication.
October 2013 Pest Report - The Netherlands. Wageningen, The Netherlands:
NPPO The Netherlands, 1.

\bibitem[Ôtake(1980)]{Otake80} Ôtake A. (1980) Chestnut gall wasp,
\emph{Dryocosmus kuriphilus} Yasumatsu (Hymenoptera: Cynipidae): a
preliminary study on trend of adult emergence and some other ecological
aspects related to the final stage of its life cycle. Appl. Entomol.
Zool. 15:96--105.

\bibitem[Piao and Moriya(1992)]{Piao92}Piao C. S., Moriya S. (1992)
Longevity and oviposition of \emph{Torymus sinensis} Kamijo and two
strains of \emph{T. beneficus} Yasumatsu et Kamijo (Hymenoptera: Torymidae).
Jap. J. Appl. Entomol. Zool. 36:113\textendash{}118.

\bibitem[Quacchia et al.(2008)]{Quacchia08}Quacchia A., Moriya S.,
Bosio G., Scapin G., Alma A. (2008) Rearing, release and settlement
prospect in Italy of \emph{Torymus sinensis}, the biological control
agent of the chestnut gall wasp \emph{Dryocosmus kuriphilus}. BioControl
53:829--839.

\bibitem[Quacchia et al.(2013)]{Quacchia13}Quacchia A., Ferracini
C., Nicholls J. A., Piazza E. Saladini M. A., Tota F., Melika G.,
Alma A. (2013) Chalcid parasitoid community associated with the invading
pest \emph{Dryocosmus kuriphilus} in north-western Italy. Insect conservation
and diversity 6(2):114--123.

\bibitem[Rieske(2007)]{Rieske07}Rieske L. K. (2007) Success of an
exotic gallmaker, \emph{Dryocosmus kuriphilus}, on chestnut in the
USA: a historical account. EPPO Bull 37:172--174.

\bibitem[Skellam(1951)]{Skellam51}Skellam, J. G. (1951) Random dispersal
in theoretical populations. Biometrika 38:196--218.

\bibitem[Vandermeer and Goldberg(2013)]{Vandermeer&Goldberg13}Vandermeer
J. H. and Goldberg D. E. (2013) Population Ecology: first principles.
Princeton University Press, II edition, Princeton (NJ), USA.

\bibitem[Viswanathan et al.(2008)]{Viswanathan08}Viswanathan G. M.,
Raposo E. P., da Luz M. G. E. (2008) Lévy flights and superdiffusion
in the context of biological encounters and random searches. Physics
of Life Reviews 5(3):133--150.\end{thebibliography}
\end{document}